\documentclass[pre,twocolumn,aps,amssymb,footinbib,showpacs]{revtex4-1}
\usepackage{amssymb}
\usepackage{amsmath}
\usepackage{times}
\usepackage{latexsym}
\usepackage{graphicx}
\usepackage{color}
\usepackage{bm}
\usepackage{subfigure}
\begin{document}
\title{ Topology and Self-Similarity of the Hofstadter Butterfly}
\author{ Indubala I Satija }
 \affiliation{School of Physics and Astronomy and Computational Sciences, George Mason University,
 Fairfax, VA 22030 }
\date{\today}
\begin{abstract}
We revisit the problem of self-similar properties of the Hofstadter butterfly spectrum, focusing on spectral as well as topological characteristics. In our studies involving any value of magnetic flux and arbitrary
flux interval,
we single out the most dominant hierarchy in the spectrum, which is found to be associated with an irrational number $\zeta=2+\sqrt{3}$ where nested set of butterflies describe a kaleidoscope.
Characterizing an intrinsic frustration at smallest energy scale, this hidden quasicrystal  
encodes hierarchical set of topological quantum numbers associated with Hall conductivity and their scaling properties. This topological hierarchy  maps to
an  {\it  integral Apollonian gasket} near-$D_3$ symmetry, revealing a hidden symmetry of the butterfly as the energy and the magnetic flux intervals shrink to zero. With a periodic drive that induces phase transitions in the system,
the fine structure of the butterfly is shown to be amplified
making states with large topological invariants  accessible experimentally.
\end{abstract}
\pacs{03.75.Ss,03.75.Mn,42.50.Lc,73.43.Nq}
\maketitle

\section{Introduction}

Hofstadter butterfly\cite{Hof}, also known as Gplot\cite{Gplot} is a fascinating two-dimensional spectral landscape, a quantum fractal
 where energy gaps
encode topological quantum numbers associated with the Hall conductivity\cite{QHE}. 
This intricate mix of order and complexity is due to frustration, induced by two competing periodicities as electrons 
in a crystalline lattice are subjected to a magnetic field.
The allowed energies of the electrons are discontinuous function of the magnetic flux penetrating the unit cell, while the gaps , the forbidden energies are continuous except at discrete points. 
The smoothness of spectral gaps in this quantum fractal may be traced to topology which makes spectral properties stable with respect to small fluctuations in the magnetic flux when Fermi energy resides in the gap.
The Gplot continues to arouse a great deal of excitement since its discovery,
and there are various recent attempts to capture this iconic spectrum in laboratory\cite{Moire,HBexp1,HBexp2}.

Fractal properties of the butterfly spectrum have been the subject of various theoretical studies\cite{Wan, Mac, Wil1, Wil2,Niu,goldman}. However,
detailed description quantifying self-similar universal properties of the butterfly and the universal fixed point butterfly function has not been reported previously.
In contrast to earlier studies where self-similarity of the spectrum is studied for a fixed value of the magnetic flux such as
the golden-mean,  and thus focusing on certain isolated local parts of the spectrum, this paper presents self-similar butterfly that is reproduced
at {\it all scales in magnetic flux}.  

In this paper, we address following questions regarding the butterfly fractal:  (1) How to describe self-similar fractal properties  of the butterfly at any value of magnetic flux given arbitrary flux interval. We determine the recursion relation, for  determining  the magnetic flux interval from one generation to the next , so that one reproduces
the entire butterfly structure . In other words, we seek the fixed point function that contains the entire Gplot as the magnetic flux and the energy scales shrinks to zero.
We try to answer this question without confining to a specific magnetic flux value such as the golden-mean and obtain universal scaling  and the fixed point butterfly function that describes the spectrum globally. (2) In addition to spectral scaling, we also address  the  question of  scaling for  the topological quantum numbers . (3) We briefly investigate butterfly fractal for special values of magnetic flux such as the golden and the silver-mean that has been the subject of almost all previous studies. (4) Finally, we present a mechanism for amplifying small gaps
 of the butterfly fractal, making them more accessible in laboratory.

Our approach here is partly geometrical and partially numerical. Using simple geometrical and number theoretical tools, we obtain the exact scaling associated with the magnetic flux interval.
Here we address the question of both magnetic flux as well as topological scaling.  The spectral gaps are labeled by two
quantum numbers which we denote as $\sigma$ and $\tau$. The integer $\sigma$ is the Chern number 
, the quantum number associated with Hall conductivity\cite{QHE} and 
$\tau$ is an integer. These quantum numbers satisfy the Diophantine equation (DE)\cite{Dana},
\begin{equation}
\rho= \phi \sigma +\tau
\label{Dio}
\end{equation}
where $\rho$ is the particle density when Fermi level is in the gap and $\phi$ denotes the magnetic flux per unit cell. We obtain exact expressions describing scaling of these quantum numbers in the butterfly hierarchy. 
The spectral scaling
describing universal scalings associated with the energy interval is obtained numerically.
Our analysis is mostly confined  to the energy scales near $E=0$, that is near half-filling. This is a reasonable choice for two reasons: firstly, simple observation of the butterfly spectrum shows gaps of the spectrum
forming $4$-wing  structures ( the butterflies)
exist mostly near half-filling. Secondly,  gaps away from half-filling appear to be continuation of the gaps that exist near $E=0$. We believe that although the gaps characterized by a fixed $(\sigma \tau)$ are discontinuous at rational values of the flux,
, these gaps continue ( with same topological numbers )  after a break  at rational flux values, with their derivatives w.r.t the magnetic flux continuous

\subsection{Summary of the main results }
\begin{itemize}

\item Given an arbitrary value of magnetic flux $\phi_0$ and an arbitrary flux interval $\delta \phi$, (no matter how small), there is a precise rule for obtaining the entire butterfly in that interval, as described in section (3-A).

\item Simple number theory provides an exact scaling ratio between two successive generations of the butterfly and this scaling is universal, independent of the initial window for zooming , described in subsections (3-B) and (IV).

\item The  hierarchy characterized by the irrational numbers whose tail exhibit period-$2$ continued fraction expansion with entries 
$1$ and $2$ ,  which we denote as $\zeta_{1,2}$ emerges as  the most "dominant" hierarchy 
as  is associated with the smallest scaling ratio that describes butterflies between two successive generations.  Commonly studied hierarchies characterized by golden-tail, which we denote as $\zeta_1$
 ( set of irrationals who tail end exhibits integer $1$ only in its continued fraction expanding) are of lower significance as they are characterized by larger scaling ratio. 
 A comparison between different hierarchies is given in Table III.
 
\item The emergence of $\zeta_{1,2}$ class of irrationals with the universal butterfly and its topological hierarchy of quantum numbers reveals a hidden dodecagonal quasicrystalline symmetry\cite{Socolar} in the butterfly spectrum.
These results also apply to other lattices such as graphene in a magnetic field.

\item  The dominant hierarchy $\zeta_{1,2}$  maps to a geometrical fractal known as the {\it  Integral Apollonian gasket}\cite{APG} that asymptotically exhibits  near $D_3$ symmetry and the nested set of butterflies describe a kaleidoscope
where two successive generations of butterfly are mirror images through a circular mirror. This is discussed in section V.
 
\item In our investigation of the fractal properties of
the Hofstadter butterfly, one of the key guiding concepts is
a corollary of the DE equation that quantifies the topology of the fine structure near rational fluxes .
We show that, for every rational flux, an
infinity of possible solutions of the DE, although not supported in the simple square lattice model
, are present in close vicinity of the flux. ( See section (IV)).
Consequently, perturbations that induce topological phase transitions can transform tiny gaps with large topological
quantum numbers into major gaps. This might facilitate the creation of such states in an experimental setting.
In section VII, we illustrate this amplification by periodically driving the system. 

\end{itemize}

\section{Model System and Topological Invariants }

Model system we study here consists of (spinless) fermions in a square lattice. 
Each site is labeled by a vector ${\bf r}=n\hat{x}+m\hat{y}$, where $n$, $m$ are
integers, $\hat{x}$ ($\hat{y}$) is the unit vector in the $x$ ($y$) direction, and $a$ is the lattice spacing. The tight binding Hamiltonian has the form
\begin{equation}
H=-J_x\sum_{\bf r}|\mathbf{r}+\hat{x} \rangle\langle \mathbf{r}|
-J_y\sum_{\bf r}|\mathbf{r}+\hat{y} \rangle e^{i2\pi n\phi} \langle \mathbf{r}|
+h.c. \label{qh}
\end{equation}
Here, $|\mathbf{r}\rangle$ is the Wannier state localized at site $\mathbf{r}$. $J_x$ ($J_y$)
is the nearest neighbor hopping along the $x$ ($y$) direction.
With a uniform magnetic field $B$ along the $z$ direction,
the flux per plaquette, in units of
the flux quantum $\Phi_0$, is $\phi=-Ba^2/\Phi_0$. 

In the Landau gauge realized in experiments\cite{Ian2}, the vector potential
$A_x=0$ and $A_y = -\phi x$, the Hamiltonian is cyclic in $y$ so
the eigenstates of the system
can be written as $\Psi_{n,m}= e^{ik_y m} \psi_n $ where $\psi_n$ satisfies the Harper equation\cite{Harper}
\begin{equation}
e^{ik_x}\psi^r_{n+1}+e^{-ik_x} \psi^r_{n-1} + 2\lambda \cos ( 2 \pi n \phi+ k_y)\psi^r_n = E \psi^r_n .
\label{harper}
\end{equation}
Here $n$ ($m$) is the site index along the $x$ ($y$) direction, $\lambda=J_y/J_x$
and $\psi^r_{n+q} =\psi^r_n$, $r=1, 2, ...q$ are linearly independent solutions.
In this gauge the magnetic Brillouin zone
is $ -\pi/qa \le k_x \le \pi/qa$ and $-\pi \le k_y \le \pi$.

At flux $\phi=p/q$, the energy spectrum has in general $q-1$ gaps.
For Fermi level inside each energy gap, the system
is in an integer quantum Hall state\cite{QHE} characterized by its Chern number
$\sigma$, the quantum number associated with the transverse conductivity $C_{xy}=\sigma {e^2}/{h}$\cite{QHE}.
The  $\sigma$ and $\tau$  are two quantum numbers that label various gaps of the butterfly and are solutions of DE\cite{Dana}. The possible values of these integers are,
\begin{equation}
(\sigma, \tau ) = (\sigma_0-n q, \tau_0 +np )
\label{DEsol}
\end{equation}
Here $\sigma_0, \tau_0$ are any two integers that satisfy the Eq. (\ref{Dio}) and $n$ is an integer.
The quantum numbers $\sigma$ that determines the quantized Hall conductivity corresponds to
the change in density of states when the magnetic flux quanta in the system 
is increased by one and whereas the quantum number $\tau$ is the change in density of states
when the period of the potential is changed so that there is one more unit cell in the system.

For any value of the magnetic flux , the system described by the Hamiltonian (\ref{qh}), supports  only $n=0$ solution  of Eq. (\ref{DEsol}) for the
quantum numbers $\sigma$ and $\tau$.  Absence of changes in topological states  from $n=0$ to higher $n$ values is
due to the absence of any gap closing and reopening that is essential for any topological phase transition.
However, as shown later, the DE which relates continuously varying quantities $\rho$ and $\phi$
with integers  $\sigma$ and $\tau$, has some important consequences about topological changes in close vicinity of rational values of $\phi$.

\begin{figure}[htbp]
\includegraphics[width =0.75\linewidth]{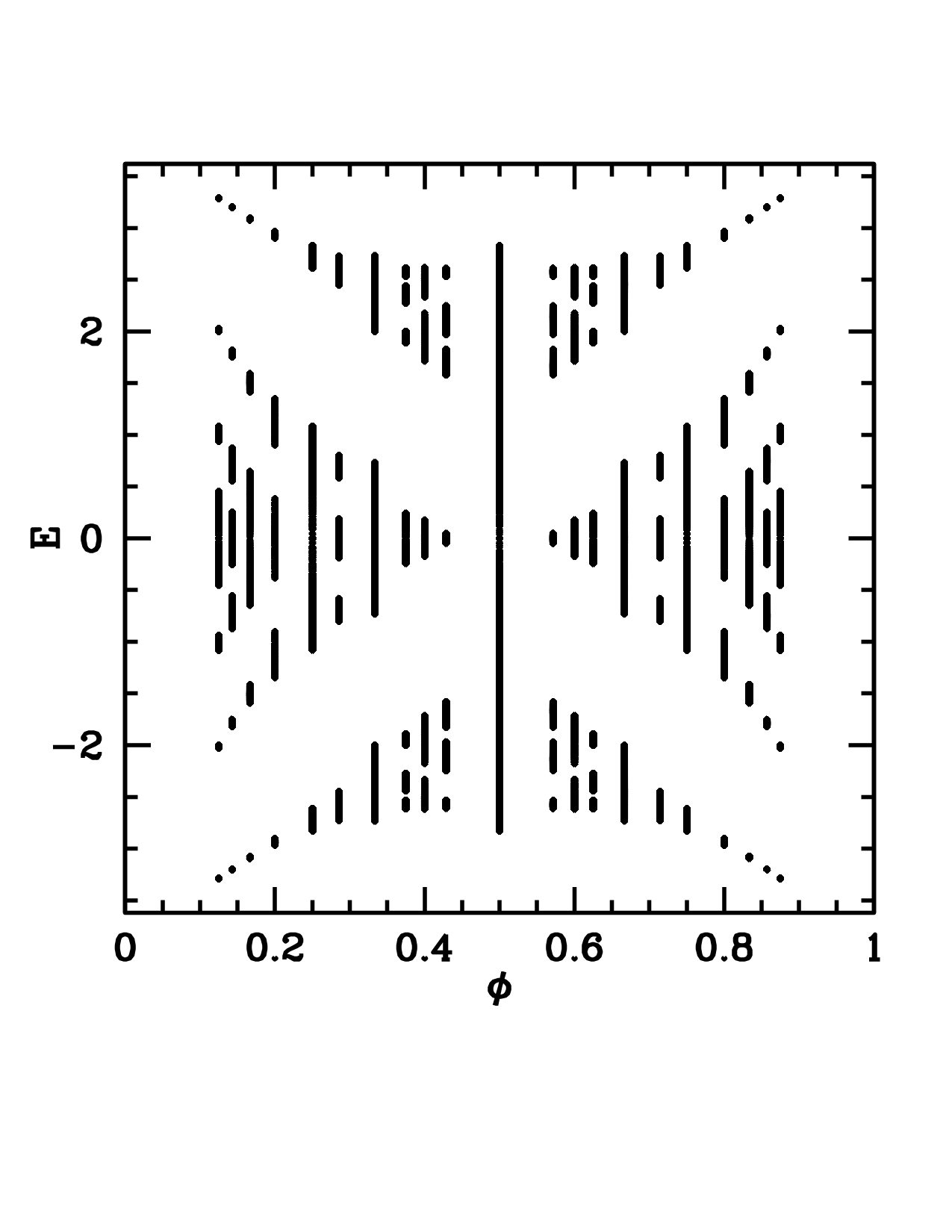}
\leavevmode \caption{ Figure shows the butterfly graph for values of $\phi=p/q$ with $q_{max}=8$.  For fixed $p/q$, the energy ( vertical axis) shows $q$ bands and $q-1$ gaps for odd-$q$ case. For $q$-even,
the two bands at the center touch and therefore there are only $q-2$ gaps.}
\label{Brational}
\end{figure}

\begin{figure}[htbp]
\includegraphics[width =0.6\linewidth]{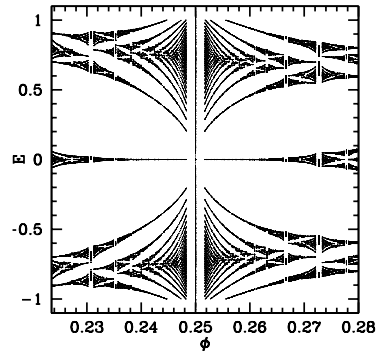}\\
\includegraphics[width =0.6\linewidth]{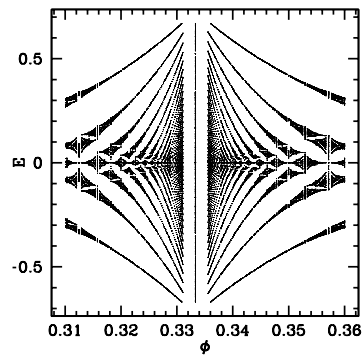}
\leavevmode \caption{ Graphs highlight the distinction between the even and the odd-denominator flux values by showing the  central band of the butterfly spectrum in the immediate neighborhood of
 $\phi=1/4$ ( left) and $\phi=1/3$ ( right). This illustrates typical scenario where near even denominator fractions the fragmented band structure clusters into two distinct bands, touching at the center while for the odd-denominator fractions, the fragmented structure clusters around a single band.}
\label{Bevenodd}
\end{figure}

\section{ Butterfly Fractal }

\subsection{Miniature Copies of the Butterfly Graph: Butterfly at Every Scale}

\begin{figure}[htbp]
\includegraphics[width = 1.1\linewidth,height=1.1\linewidth]{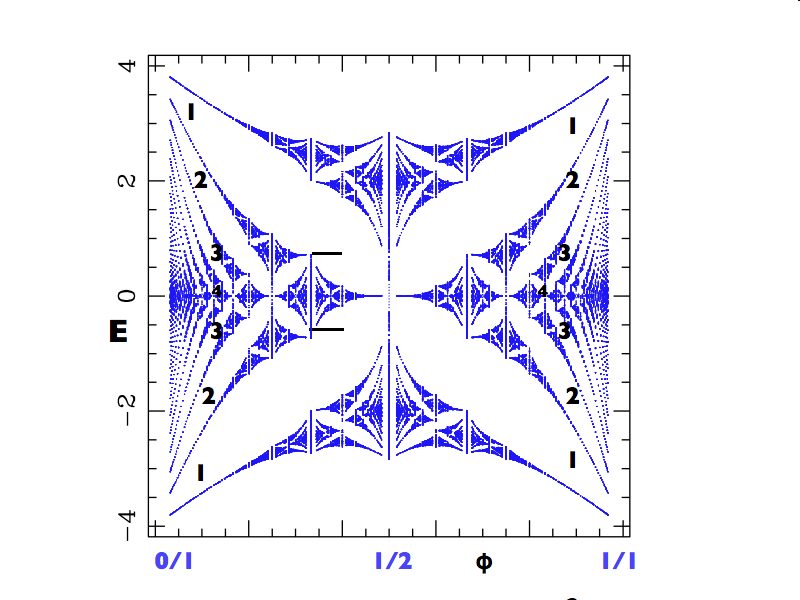}
\leavevmode \caption{(color on line) 
 Level-$1$ butterfly in $\phi$ intervals $[1/3-2/5]$  where the horizontal bars show the $\phi$-interval that is zoomed in the level-$2$.
 The magnitude of the Chern numbers for
 the central butterfly and its left and right harmonics are $1+3n$ and $|1-5n|$, $n=1,2,3..$ as shown are determined using DE equation.}  
\label{butt0}
\end{figure}

\begin{figure}[htbp]
\includegraphics[width = 1.1\linewidth,height=1.1\linewidth]{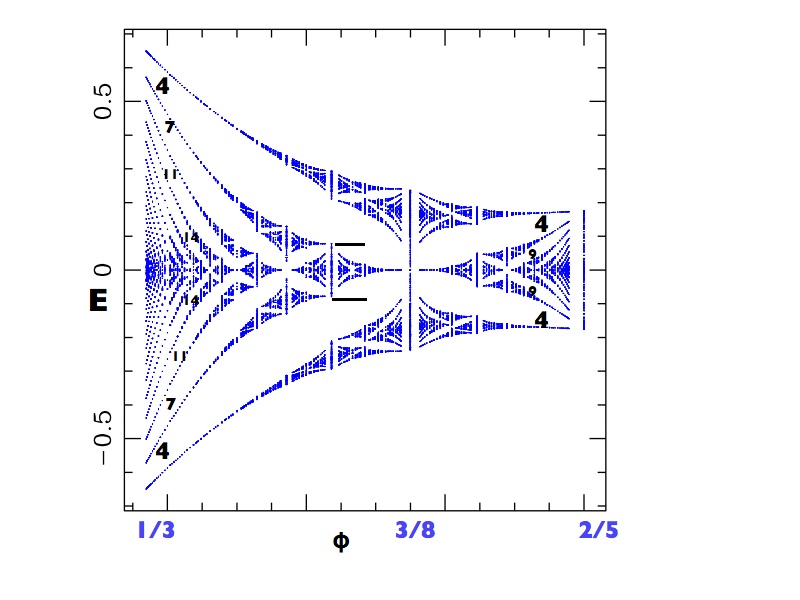}
\leavevmode \caption{(color on line) 
 Level-$1$ butterfly in $\phi$ intervals $[1/3-2/5]$  where the horizontal bars show the $\phi$-interval that is zoomed in the level-$2$.
 The magnitude of the Chern numbers for
 the central butterfly and its left and right harmonics are $1+3n$ and $|1-5n|$, $n=1,2,3..$ as shown. } 
\label{butt1}
\end{figure}

\begin{figure}[htbp]
\includegraphics[width = 1.0\linewidth,height=1.0\linewidth]{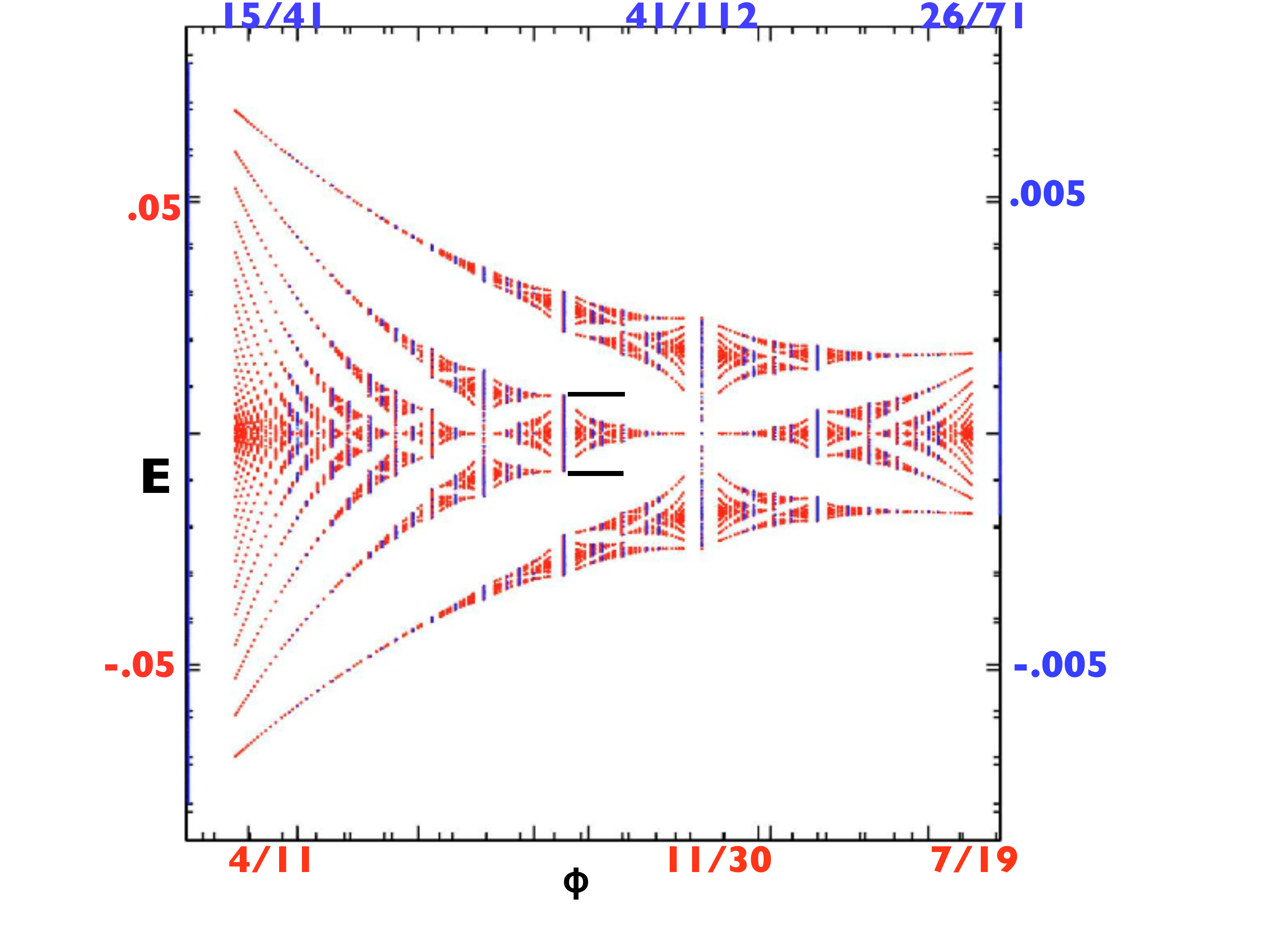}
\leavevmode \caption{(color on line) 
Figure displays an overlay of level-$2$ and level $3$ butterflies in $\phi$ intervals $[4/11,7/19]$   and
$[15/41,26/71]$ giving energy scale ratio $\approx 10$ as shown in red and blue respectively. The magnetic flux scaling is found to be $\zeta^2$ as described later.
 This illustrates an approach to fixed point butterfly fractal as $E$ and $\phi$ intervals shrink to zero. Note that the $\phi$ scale is shown only at the boundaries and the centers of the butterflies.
 The magnitude of topological numbers for the central butterflies for level-$2$ and $3$ are respectively given by $(15,3)$ and $(56,20)$. The asymmetry of the butterfly  about its center reflects the asymmetry in the Chern numbers of the left and the right harmonics which can be determined from Eq. (\ref{harmonics}). }
\label{UB}
\end{figure}

Butterfly graph is a plot of possible energies of the electron for various values of $\phi$ which varies between $[0,1]$.  To understand this graph, we begin with
values of $\phi$ that are rational numbers, focusing on simple fractions. Figure (\ref{Brational}) shows  one such graph, a
skeleton of the butterfly graph, obtained using few rational values. The permissible energies are arranged in {\it bands} separated from forbidden values, the {\it gaps}.
In general, for a fixed  $\phi=\frac{p}{q}$,  the graph consists of
of $q$ bands ( the dark regions) and $(q-1)$ gaps ( empty regions).\\
 
For $q$-even, the two central bands touch one another and therefore, we see only $(q-2)$ gaps. The graphs show important distinctions between the even and the odd-denominator  fractions as shown in Fig.(\ref{Bevenodd}).
As we look in the immediate vicinity of even-denominator flux values, 
 the two touching bands begin to split, opening a gap at the center.
Consequently, in the butterfly landscape,  as we look both to the left and to the right of the even-denominator fraction, we see  four gaps or swaths resembling the four-wings of a butterfly, all converging at the center. 
In contrast, near odd-denominator fractions, we  see a proliferation of a set of discrete levels,  that cluster around a single band, namely the central band corresponding to the odd-denominator fractions.\\

Therefore, every even-denominator fractional flux value forms the center of a butterfly .To find miniature butterfly ( centered at $E=0$) in the butterfly graph near an arbitrary location in $\phi=\phi_0$ and with a scale, say  $\delta \phi$, 
\begin{itemize}

\item Pick any irreducible fraction , say $f_c=\frac{p_c}{q_c} \approx \phi_0$,  where $q_c$ is even and $(q_c)^2 \approx  \delta \phi$.

\item In a Farey sequence $F_{q_c}$ ( sequence that consists of all irreducible rationals with $q_{max}=q_c$), locate the left and right Farey neighbors of $f_c$ which we denote as 
$f_L=\frac{p_L}{q_L}$ and  $f_R=\frac{p_R}{q_R}$.  Simple number theoretical reasoning shows that for every given $f_c$, there is a unique pair $f_L$ and $f_R$  which are Farey neighbors of $f_c$.  ( See Appendix )

\item Determine the widths of the central band ( located symmetrically about $E=0$ ),  corresponding to fractions $f_L$ and $f_R$, denoted as $\Delta E_L$ and $\Delta E_R$, by diagonalizing the Harper equation.

\item The  miniature butterfly is sub-part of the butterfly graph, symmetrically located about $E=0$ with butterfly center at $(\phi=f_c, E=0)$ and its left and right boundaries confined between
$(f_L, \pm \frac{\Delta E_L}{2})$, and  $(f_R, \pm \frac{\Delta E_R}{2})$. In other words, near any even-denominator fraction, one can find a unique butterfly, with flux interval $\Delta \phi_c = (f_R-f_L)=\frac{1}{q_L q_R}$ 
( horizontal scale) and bounded vertically by $\Delta E_L$ and $\Delta E_R$
on the left and on the right respectively.

\item Since $f_c$ is a Farey neighbors of both $f_L$ and $f_R$, the Ford circles representing these fractions touch and such butterflies satisfy the condition, $f_c = f_L \bigoplus f_R$. We note that for a 
fixed $f_c$, $f_L$ and $f_R$, the Farey neighbors of $f_c$ are uniquely determined and there is no additional butterfly in the interval $\Delta \phi_c$ for $q < q_c$. 
\end{itemize}

Table I illustrates the process of deterring the boundaries of the butterfly, once we choose its center.
\\

At each level $l$, we label the rational flux values at the center, the left and the right boundaries as
$f_c(l)=\frac{p_c(l)}{q_c(l)}$,   $f_L(l)=\frac{p_L(l)}{q_L(l)}$ and $f_R(l)=\frac{p_R(l)}{q_R(l)}$ respectively. 

\begin{widetext}

\begin{table}
\begin{tabular}{| c | c |  c | c |}
\hline
 $f_c$\,\, & Farey Sequence Needed & $f_L$& $f_R$\\
 \hline
$1/4$ & $F_4$: ${\bf 0/1,  1/4, 1/3}$, $1/2$& $0/1$ &$1/3$\\
\hline
$1/6$  &$F_6$: ${\bf 0/1,  1/6, 1/5}$, $1/4, 1/3, 2/5, 1/2$, &$0/1$,& $1/5$\\
\hline
$3/8$ &$F_8$:  $0/1, 1/8,1/7, 1/6, 1/5, 1/4,  2/7$, ${\bf 1/3, 3/8,  2/5}$,  $3/7, 1/2$, &$1/3$ &$2/5$\\
\hline
$1/8$ & $F_8$: ${\bf 0/1,  1/8, 1/7}$, $1/6, 1/5, 1/4,  2/7,1/3, 3/8, 2/5,  3/7, 1/2$&  $0/1$ &$1/7$\\
\end{tabular}
\caption{ Given the center, locating the left and the right boundaries of the butterfly where the center and the boundaries are shown in bold.}\label{table1}
\end{table}

\end{widetext}

Figures ~ \ref{butt0} and (\ref{butt1}) and (\ref{UB})  show  numerically obtained energy spectrum displaying four successive blowups of butterfly structures.

To describe hierarchical structure of the butterfly fractal, we introduce a notion of "levels"  (or generations), where higher levels  ( generations) correspond to viewing the butterfly at smaller and smaller scale
in $E$ vs $\phi$ plot. At level-$0$ we have the central butterfly in the $\phi$-interval $[0,1]$ with center at $\phi_c=1/2$ and
 colonies of butterflies to the left as well as to the right of $\phi=1/2$.  ( See Fig. (\ref{butt0}) The left colony, all sharing a common left boundary at $\phi=0$ are centered at $\frac{1}{2n}$ .(
Similarly, there is a right colony, centered at $1-\frac{1}{2n}$, all sharing the right boundary at $\phi=1$.
Therefore, the boundaries of the central butterfly enclose the boundaries of the  left and the right colony.  

{\it When we refer to a {\it Butterfly} in the Gplot, we mean a central butterfly and a  set of left and a set of right colonies of butterfly that share respectively the left and the right boundary of the central butterfly as  discussed below,
this entire structure is reproduced at all length scales.} 

The level-$l+1$ butterfly resides in a smaller flux interval that is  entirely contained in the flux-$l$ interval
of the level-$l$ butterfly.  In other words, neither the left nor the right boundary points of level-$l+1$ overlap with the boundaries of the level-$l$.
We note that beyond level-$0$, butterflies do not exhibit reflection symmetry about their centers.

Fig. (\ref{UB}) suggests the existence of a fixed point butterfly as two successive levels overlay. We note that  choice of  any magnetic flux interval in the butterfly fractal leads to similar result as discussed later in the paper.

\subsection{ Recursion Relations  for Magnetic Flux Interval }

We will now describe the scaling of the magnetic flux intervals as one zooms into the butterfly fractal.\\

A close inspection of the Gplot reveals that Farey sequences are the key to systematically sub-divide the $\phi$ interval, where each new interval reproduces the entire butterfly.
By {\it Farey path, we mean a path in the Farey tree that leads from level-$l$ to level $l+1$, connecting the centers of the butterfly at 
two successive levels, through its boundaries}.  We want to emphasize that our reference to "Farey tree" does not correspond to a path that connects rational approximants
of an irrational number,  it is a path that finds the entire butterfly ( its boundaries and center ) between two generations or levels of hierarchy. 
This Farey path described for various different parts of the Gplot , is encoded in the following recursive set of equations,

\begin{eqnarray}
f_L(l+1)&=&f_L(l) \bigoplus f_c(l)\nonumber\\
f_R(l+1)&=&f_L(l+1) \bigoplus f_c(l)\nonumber\\
f_c(l+1)&=&f_L(l+1) \bigoplus f_R(l+1),
\label{RR}
\end{eqnarray}

where the Farey sum, denoted by $\bigoplus$ between two rationals $\frac{p1}{q1}$ and $\frac{p2}{q2}$ is defined 
as $\frac{p1}{q1} \bigoplus \frac{p2}{q2}=\frac{p1+p2}{q1+q2}$. Since $f_L$ and $f_R$ are neighbors in the Farey tree ( see Appendix ), we have,

\begin{equation}
p_R q_L - p_L q_R = 1
\label{neighbor}
\end{equation}

Simple calculations lead to following recursion relations for $p_x$ and $q_x$ where $x=c,L,R$:

\begin{equation}
p_x(l+1)=4p_x(l)-p_x(l-1), \quad q_x(l+1)=4q_x(l)-q_x(l-1)
\label{pqRR}
\end{equation}

We now define the ratio $\zeta(l) = \frac{q_x(l+1)}{q_x(l)}$ and Eq. (\ref{pqRR}) gives,
\begin{equation}
\zeta(l) = 4 -\frac{1}{\zeta(l-1)},
\end{equation}

We now define $\zeta=  \lim_{ l \rightarrow \infty} \zeta(l)$, where $\zeta$ satisfies the following equation,

\begin{equation}
(\zeta)^2-4\zeta+1=0, \quad \zeta = 2+\sqrt{3}
\label{Rfix}
\end{equation}

We can now calculate the scaling associated with the magnetic flux, the horizontal scale of the butterfly. At a given level $l$,  the magnetic flux interval that contains the entire butterfly is,

\begin{equation}
\Delta \phi(l)=f_R(l) - f_L(l) = \frac{1}{q_L(l) q_R(l)}
\label{dphi}
\end{equation}
Therefore, we obtain the scaling associated with $\phi$, which we denote as $R_{\phi}$,

\begin{equation}
R_{\phi} = \lim_{ l \rightarrow \infty} \frac{\Delta \phi(l)}{\Delta \phi(l+1)} = \zeta^2
\end{equation}

To calculate $\Delta(l)$ for a an arbitrary level-$l$, we use Ford circles ( see the Appendix) that provide a pictorial representations of fractions.

\begin{table}
\begin{tabular}{| c | c |  c | c| }
\hline
 $\Delta \phi(1)$\,\, & $\sigma$  & $\tau=\frac{\beta-1}{2}$ & $\phi^*_c$ \\ \hline
 $ [2/5-1/3] $\,\,&  $(4,15,56,209..)$ &  $(1, 5, 20, 76..)$ & $[2, 1, 2, 1..] =\frac{\sqrt{3}-1}{2}$
 \\ \hline
  $[1/3-2/7]$\,\, & $( 5, 18, 67, 250..)$ & $( 1, 5, 20, 76..)$& $[3,3,1,2,1,2...]$
  \\ \hline
   $ [2/9-1/5] $\,\,&  $(7,26,97,362..)$ &  $(1, 5, 20, 76..)$ & $[4,1,2,1,2,1,2..] $
 \\ \hline
  $[3/7-2/5]$\,\, & $( 6,23,86,321....)$ & $( 2,9,35,132..)$& $[2,2,2,1,2,1,2...]$\\ \hline
\end{tabular}
\caption{ Topological integers and the centers of the fixed point butterflies for four \underline{initial} flux intervals, $\Delta \phi(1)$,
and following the Farey path "LRL" between two successive levels. In all cases, the centers of the fixed point butterflies are found
to be irrational numbers whose tails exhibit
periodic continued fraction with entries $1$ and $2$. Note that the sequence of topological integers shown here correspond to  rational approximants of $\phi^*_c$. In all cases, ratio of the topological numbers converge to $R_{\sigma}=R_{\tau}$.} 
\label{table2}
\end{table}

 These solutions describe
the fine structure of the butterfly near $\phi_0$.
Consequently, the spectral gaps near $\phi=1/q$ have
Chern numbers changing by a multiple of $q$. This
suggests a semiclassical picture near $\phi=p/q$ in terms of an effective Landau level
theory with cyclotron frequency renormalized by $q$. This is demonstrated in the Fig. (\ref{butt0}) near $\phi=0$ and in  Fig.(\ref{butt1}) near $\phi=1/3$ and $2/5$.

\section{ Topological Characterization of the Butterfly }

We now calculate the quantum numbers ($\sigma \tau$) associated with various gaps of the butterfly structure at all scales.
These results are  consequence of Diophantine equation, and are based on two corollaries, $C_1$ and $C_2$ as described below.\\

$[{\bf  C_1}]$ The Chern  number associated with the four  dominant gaps that form the central butterfly  are, $\sigma_c=\pm  \frac{q_c}{2}$.\\

$[{\bf  C_2}]$  Chern numbers  of a set of gaps that begin near the boundary are given by $\sigma_b = \sigma_c \pm  n q_b$ where $q_b$ is the denominator of the fractional flux at the boundary. \\

Proof $C_1$ : $f_L$ and $f_R$ are respectively the left and the right neighbors of $f_c$ and therefore,
\begin{eqnarray}
p_L q_c -p_c q_L & = & -1\\
p_R q_c -p_c q_R & = & 1
\label{neighbor1}
\end{eqnarray}

This implies that for any integer $n \ge 0 $, $\frac{p_L+np_c}{q_L+nq_c}$ are a set of left neighbors of $f_c$ and similarly
$\frac{p_R+np_c}{q_R+nq_c}$ are a set of right neighbors of $f_c$ in the Farey tree as,

\begin{equation}
|p_c(q_x+n q_c)-q_c(p_x+ n p_c)|=1; \,\,\, x = L, R.
\end{equation}

We now calculate the Chern number near half filling for the neighbors $\frac{p_x+n p_c}{q_x+n q_c}$ of $f_c$, . This will correspond to $r=(q_x+n q_c)/2-1$.
Substituting in the DE equation, we obtain, 
\begin{equation}
\sigma = \pm \frac{q_c}{2}; \,\,\,\, \tau = \frac{1 \pm p_c}{2}
\label{TPint}
\end{equation}

We note that the central butterfly, characterized by four wings ( gaps ) is characterized by a unique pair of topological integers determined by the Eq. \ref{TPint} \\

Proof $C_2$ : Chern numbers of the set of gaps near the boundary are  given by
the infinity of solutions
depicted in Eq.(\ref{DEsol}) reside in close proximity to the flux $\phi$  and label the fine structure of the butterfly in Gplot .\\

DE equations at $\phi_0$, $\rho_0$ and in its vicinity $(\phi_0+\delta \phi)$, $(\rho_0+\delta \rho)$ are,
\begin{equation}
\rho_0= \phi_0 \sigma_0 +\tau_0
\label{E1}
\end{equation}
\begin{equation}
\rho_0+\delta \rho= (\phi_0+\delta \phi) ( \sigma_0+\Delta \sigma) +(\tau_0+\Delta)\\
\label{E4}
\end{equation}
Keeping terms linear in $\delta \rho$ and $\delta \phi$, we get
\begin{equation}
\rho_0+\delta \rho = \phi_0 \sigma_0 +\delta \phi \sigma_0+ \Delta \sigma \phi_0 + \tau_0+\Delta \tau
\label{E2}
\end{equation}
Using (\ref{E1}), Eq. (\ref{E2}) reduces to,
\begin{equation}
\delta \rho = \delta \phi \sigma_0+ \Delta \sigma \phi_0 +\Delta \tau
\label{E3}
\end{equation}
Key observation from Eq. (\ref{E3}) is that unlike $\delta \rho$ and $\delta \phi$ which are can chosen to be infinitesimally small, $\Delta \sigma$ and $\Delta \tau$ are integers and therefore,
for small $\delta \rho$ and $\delta \phi$ we get,
\begin{equation}
\phi_0 \Delta \sigma+ \Delta \tau = 0;\,\,\,\,
\frac{\Delta \sigma}{\Delta \tau}= -\frac{q_0}{p_0}
\label{Dchern}
\end{equation}

Since both $\Delta \sigma$ and $\Delta \tau$ are integers and $p_0$ and $q_0$ are relatively prime, the simplest solutions of Eq. (\ref{Dchern})
are,
\begin{equation}
 \Delta \sigma = \pm n q_0;\,\,\,\, \Delta \tau = \mp n p_0;\, \,\,n=0,1,2,....
 \label{harmonics}
 \end{equation}
 
\underline{{\it Topological Scaling}}\\

Equation (\ref{TPint}) relating the denominators of the fraction and Equations from Chapter II that gives recursions from the numerator and denominator of the fractions,  lead to the  following recursion relations for topological integers,

\begin{eqnarray}
\sigma(l+1)=4\sigma(l)-\sigma(l-1)\\
 \beta(l+1)=4\beta(l)-\beta(l-1)
\label {RLt}
\end{eqnarray}

The Eq. (\ref{RLt}) results in fixed point solution of the ratio of integers at two successive levels,\\
\begin{eqnarray}
 \frac{\sigma(l+1)}{\sigma(l)}&=&R_{\sigma}(l)=  4-\frac{1}{R_{\sigma}(l-1)}\\
  \frac{\beta(l+1)}{\beta(l)}&=&R_{\beta}(l)=  4-\frac{1}{R_{\beta}(l-1)}\\
 R_{\sigma} & = & \lim_{l \rightarrow \infty} R_{\sigma}(l) =2+\sqrt{3}\\
   R_{\beta} & = &\lim_{l \rightarrow \infty} R_{\beta}(l) =2+\sqrt{3}
 \label{fixedpt}
 \end{eqnarray}


The irrational number  $\zeta$ has a continued fraction expansion,$ \zeta=[3,1,2,1,2,1,2....]$, given by,
\begin{equation}
 \zeta = \cfrac{1}{3
          + \cfrac{1}{1
          +\cfrac{1}{2
          + \cfrac{1}{1 + \cfrac{1}{2.....} } } }}
          \label{cont}
\end{equation}
 
We will refer this irrational number as {\it diamond mean}. It is instructive to consider a somewhat general case where a butterfly 
have left (right)  boundary located at $\phi_{L(R)}= \frac{1}{(2n+1)}$. The fixed points of the centers of these butterfly and their corresponding Chern numbers are given by,
\begin{eqnarray}
\phi^*_{c_{R(L)}} &= &\frac{1}{2[n+\alpha_{L(R)}]}\\
\alpha_{R} &= &\frac{\sqrt{3}-1}{2}=\frac{1-\zeta}{2}\\
\alpha_{L} &=& 1-\alpha_R = \frac{1+\zeta}{2}\\
\sigma_R(l) & = & 1+3n\\
\sigma_L(l) & = & 2+3n
\end{eqnarray}

This illustrates the asymmetry of the universal butterfly
as the gaps on the right have smaller Chern numbers  compared to the gaps on the left. It is interesting to note that  unlike $\sigma$, the quantum number $\tau$  are same for the left and the right colonies of butterfly. We emphasize that although the topological numbers
depend upon the initial interval, the topological scaling ratio converges to the same universal value.

Asymptotically, $\sigma(l) \rightarrow \zeta^{-l}$, $\tau \rightarrow \zeta^{-l}$ 
and the underlying $\phi$ interval
scales as, $\Delta \phi \rightarrow \zeta^{2l}$.

For the butterfly fractal shown in Fig. (\ref{butt0},\ref{butt1}), the entire band spectrum is numerically found to scale approximately as,
$\Delta E(l)  \approx 10^{-l} $.
Although the precise value of quantum numbers ( and hence the universal butterfly fractals) depend upon $\phi$, the scaling ratios between two successive levels
is $\phi$ independent.

Comparing scaling exponents for the size of the butterfly, ( described by $\Delta \phi$ and $\Delta E$ ) and the corresponding 
topological quantum numbers, we note that the topological variations occur at a slower rate than the corresponding spectral variations as one views the butterfly at a smaller and smaller scale.
\\

\section{Integral Apollonian Gasket and the Butterfly Topology}

\begin{widetext}

\begin{figure}[htbp]
\includegraphics[width = .7\linewidth,height=.5\linewidth]{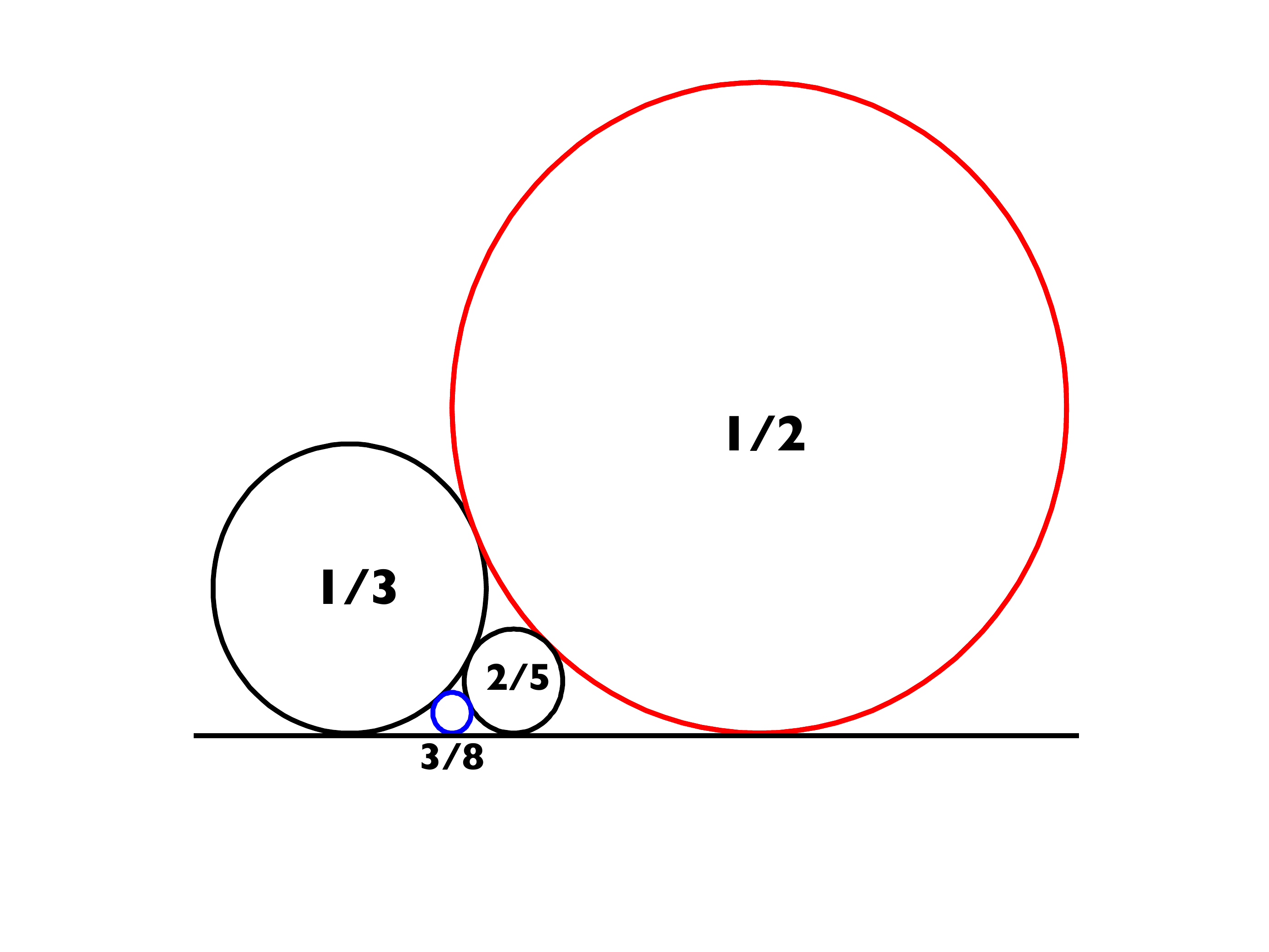}\\
\includegraphics[width =.8\linewidth]{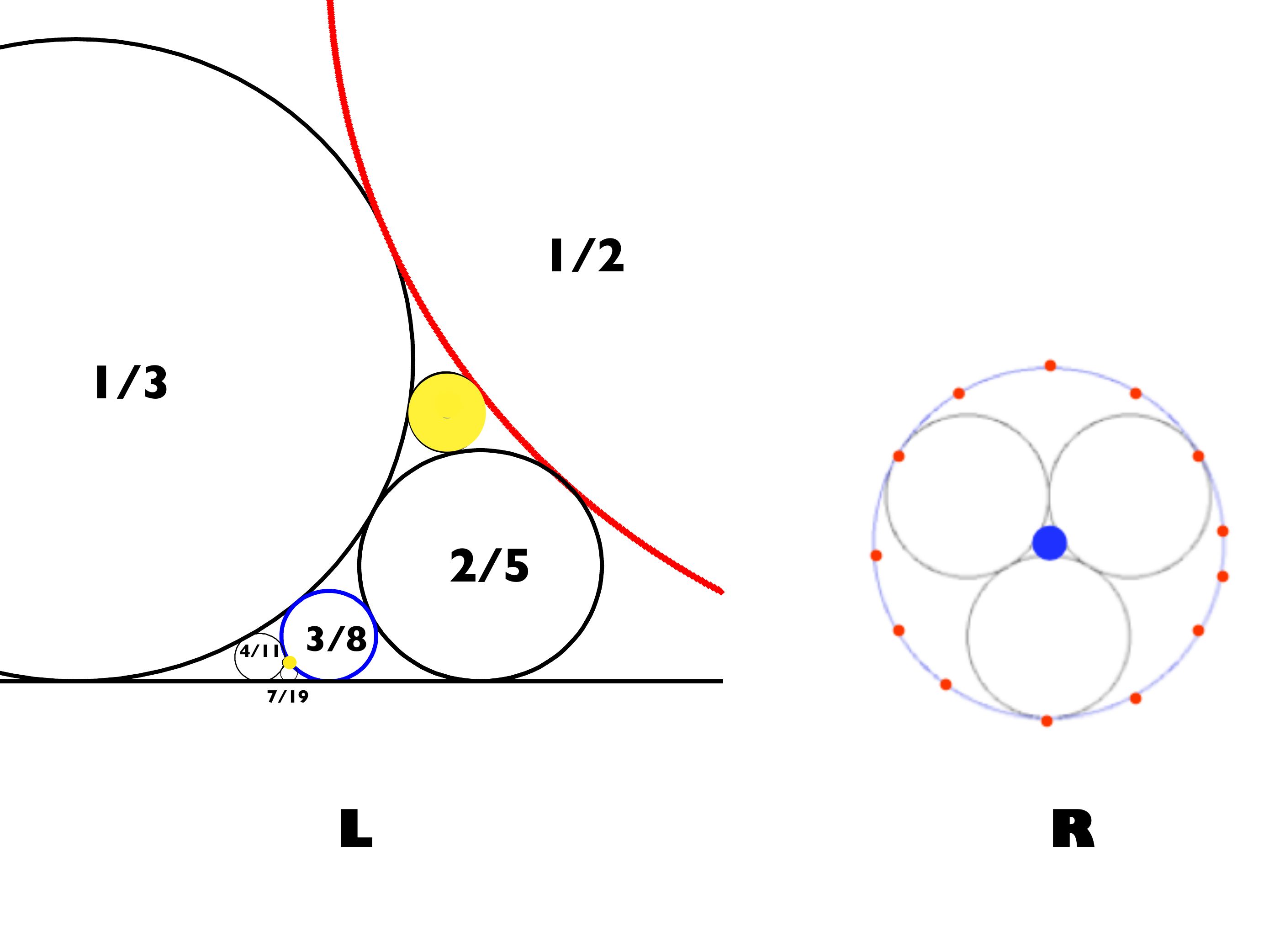}
\leavevmode \caption{ Upper Figure shows the Ford circle representation of the center ($3/8$ blue) and the left ($1/3$ black) and the right ( $2/5$ black). These three circles along with the horizontal black line
are mutually tangent .  We note that the scaling ratio between the red and the blue circles, corresponding to two consecutive generations of the butterfly asymptotically equals the ratio
of the curvatures of the innermost ( solid blue)  and the outermost  circles ( with red dots) of the Apollonian gaskets with near $D_3$ symmetry ( Lower right). 
Lower left is the blowup of the upper figure with additional level of the hierarchy. The yellow circles ( big and small) are respectively 
the image of the horizontal axis reflected through the tangency points of the circles corresponding to $1/3, 2/5, 1/2$  
and  $4/11, 7/19, 3/8$. We note that these two consecutive image circles also scale by the same ratio $(\zeta^*)^2$. }
\label{A3}
\end{figure}

\end{widetext}

 \begin{figure}[htbp]
\includegraphics[width = 1\linewidth,height=.8\linewidth]{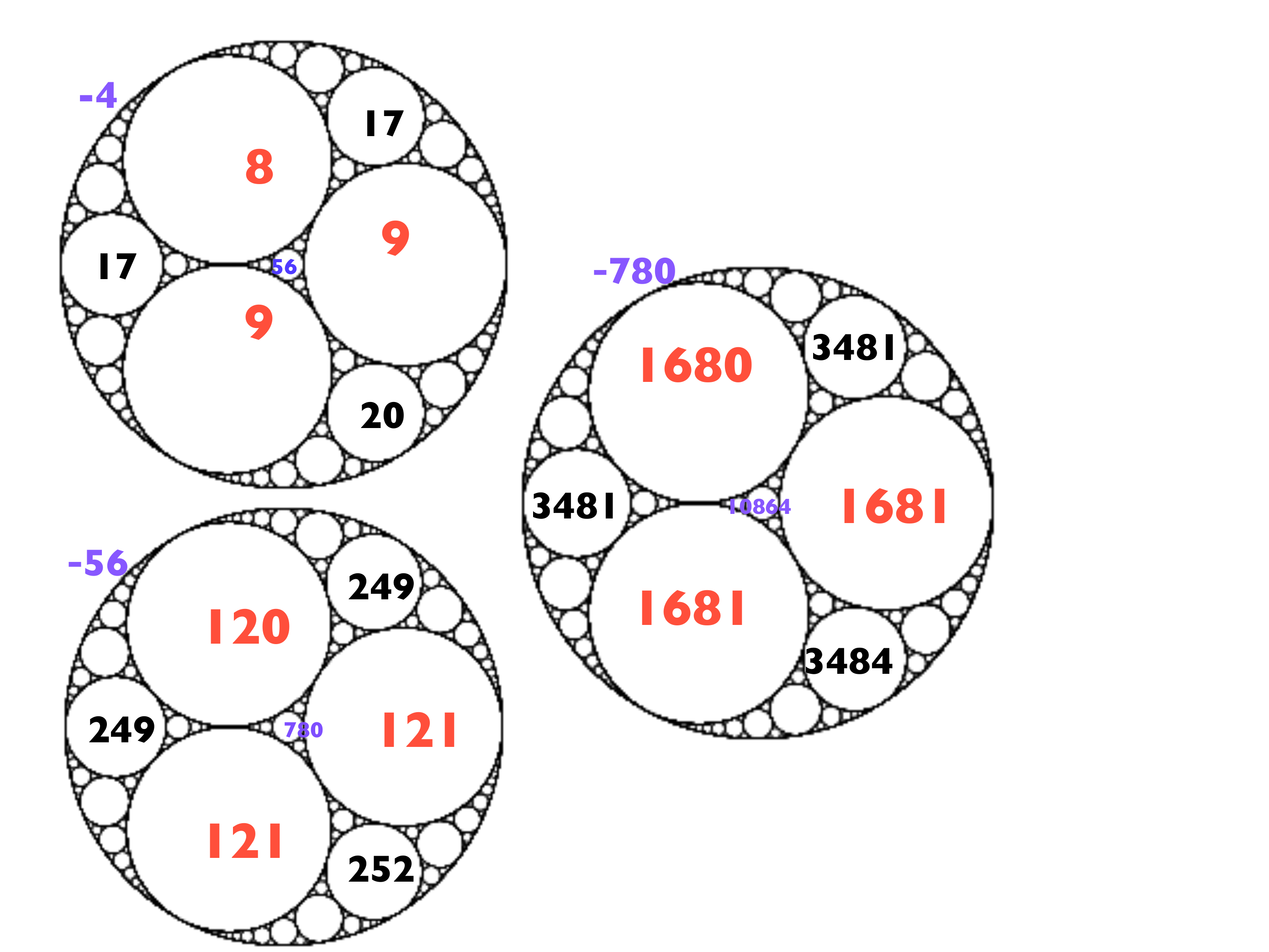}\\
\includegraphics[width = 1\linewidth,height=.8\linewidth]{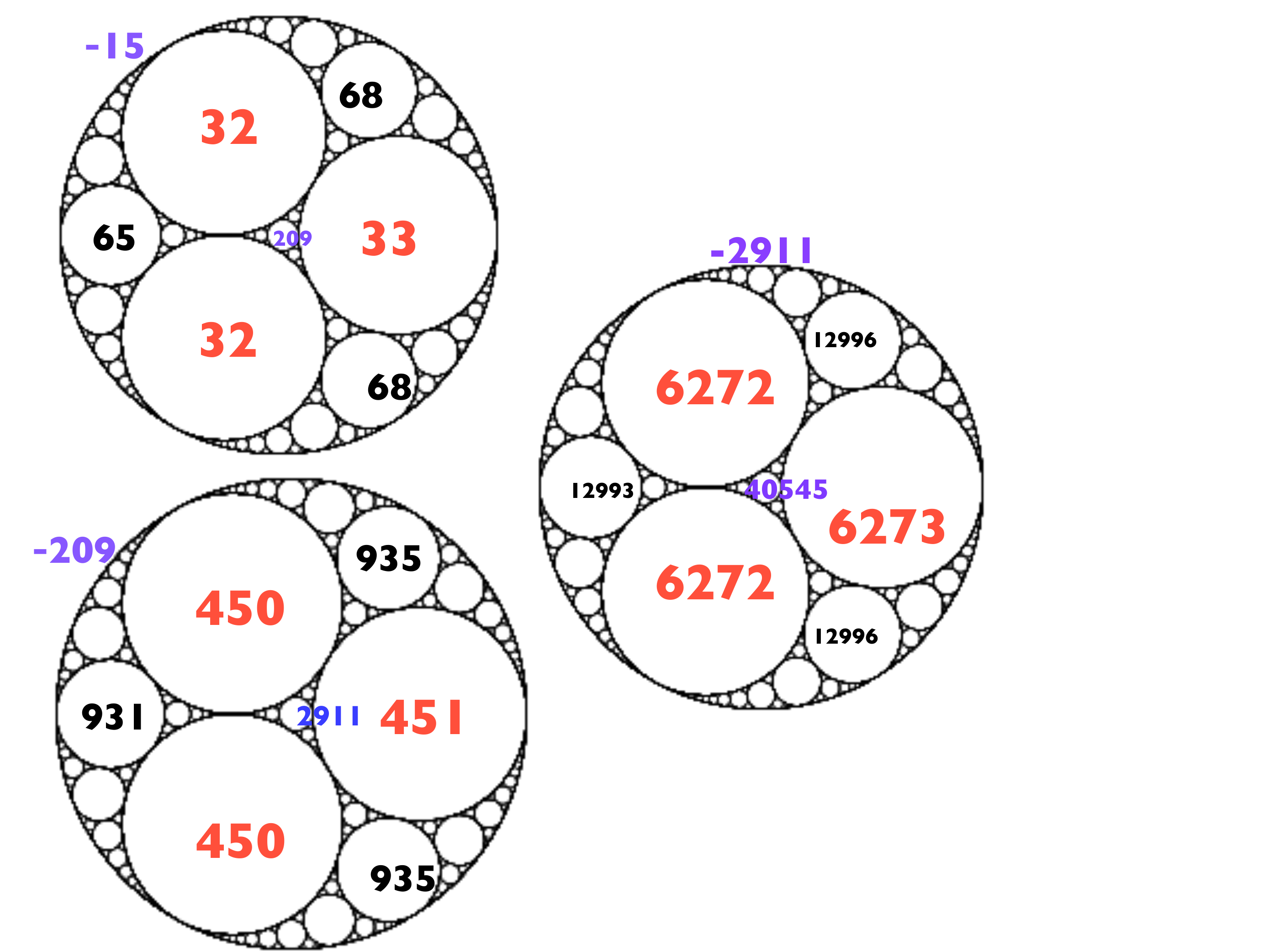}
\leavevmode \caption{ Circles show sequential construction of Apollonian gaskets where the upper (lower) three circles show three generations beginning with 
$(-4,8,9,9)$ ( $(-15,32,32,33)$ ) corresponding to even-$n$ ( odd-$n$) case. Note that the bounding circles with negative curvatures
 encode the chern numbers shown in the first row of Table I.  Asymptotically, the Apollonian gasket exhibits $D_3$ symmetry.  Also, note that all equivalent circles, asymptotically,
scale by the same factor $\zeta^2$ as shown by converging ratios in the figure and hence show the flux scaling intervals of the butterflies.}
\label{APG}
\end{figure}

The topological and magnetic flux scaling of the butterfly fractal is related to  Apollonian gaskets.
The Appendix provides a brief introduction to these fractals and discusses its
 properties that are key to understand the relationship between the butterfly fractal and the Apollonian gaskets. \\
 
The central concept that links Apollonian gaskets and the butterfly is hidden in the  pictorial representations of fractions using Ford circles. We now discuss this relationship. Brief introduction to Ford  circles is given
in the Appendix.\\

\subsection{Ford Circles, Apollonian Gasket and the Butterfly}

As stated in the Appendix,  a fraction $f=\frac{p}{q}$  can be represented by a circle ( Ford circle ) , with curvature ( inverse radius)  $\kappa= 2 q^2$. These Ford circles also provide a pictorial
representation of the size of the magnetic flux intervals. Using the Eq. (\ref{pqRR}),

\begin{equation}
\sqrt{\kappa_c(l+1)}= 4 \sqrt{\kappa_c(l)}-\sqrt{\kappa_c(l-1)}
\end{equation}

The Ford circles do not touch and are all tangent to the horizontal axis of the butterfly graph. Introducing a scale factor $\zeta(l)$ as,
\begin{equation}
 \zeta(l)=\sqrt{\frac{\kappa_c(l+1)}{\kappa_c(l)}},
 \end{equation}
we obtain,

\begin{equation}
\zeta(l)= 4-\frac{1}{\zeta(l-1)}
\end{equation}

For large $l$, $\zeta(l) \rightarrow \zeta(l+1)$, which we denote as $\zeta^*$, which satisfies the quadratic equation,

\begin{equation}
(\zeta^*)^2 -4\zeta^*+1=0, \quad \zeta^* = \lim_{ l \rightarrow \infty} \sqrt{\frac{\kappa_c(l+1)}{\kappa_c(l)}}= 2 + \sqrt{3}
\label{zetascale}
\end{equation}

Therefore, Ford circles corresponding to even-denominator fractions  form a self-similar fractal consisting of circles whose curvatures scale asymptotically by $\zeta^*$. Interestingly,
starting with different even-denominator fraction, we get a different set, all exhibiting the same scaling. We note that,\\
\begin{equation}
\sqrt{\frac{\kappa_c(l)}{\kappa_L(l)}}  \rightarrow  1+\sqrt{3},\quad 
\sqrt{\frac{\kappa_c(l)}{\kappa_R(l)}}  \rightarrow \frac{ 1+\sqrt{3}}{\sqrt{3}},\quad
\sqrt{\frac{\kappa_R(l)}{\kappa_L(l)} } \rightarrow  \sqrt{3}
\end{equation}

Figure (\ref{A3}) shows the Ford circle representation of the levels $0$ and $1$ of the butterfly centers $f_c$ and the boundaries, $f_L, f_R$. 
In general, any two successive  levels $l$ and $l+1$ of the butterfly flux intervals, we have \underline{ two} configurations  which we list below,
of four mutually tangent circles  with curvatures $k_i, i=1-4$ representing fractions\\

 (1) $f_c(l), f_L(l+1), f_R(l+1)$  and base line are mutually tangent  where $k_1 \equiv k_c(l)=2q_c^2(l)$,  $k_2 \equiv k_L(l+1)=2q_L^2(l+1)$, $k_3 \equiv  k_R(l+1) = 2q_R(l+1)^2$ and the base line with $k_4 =0$ and \\
 
  (2) $f_c(l+1), f_L(l+1), f_R(l+1)$ are mutually tangent  where $k_1 \equiv k_c(l)=2q_c^2(l)$,  $k_2 \equiv k_L(l+1)=2q_L^2(l+1)$, $k_3 \equiv  k_R(l+1) = 2q_R(l+1)^2$ and the base line with $k_4 =0$ \\

From Descartes's theorem ( Eq. (\ref{DT})), we obtain\\

\begin{equation}
k_c(\pm) = k_R(l+1) \pm k_L(l+1), 
\end{equation}

where we can identify,

\begin{equation}
k_c(+)=k_c(l+1), \quad k_c(-) = k_c(l)
\end{equation}

These two configurations describing butterfly fractal at two successive generations are in fact mirror image of each other, through a circle drawn from the tangency point of $f_L(l+1)$, $f_R(l+1)$ and the base line. In other words, the circles  with
curvatures $k_c(l)$ and $k_c(l+1)$ play the same role as the outermost and innermost circles  of the Apollonian gasket in the configuration described on the right in Fig (\ref{A3}). \\

To see explicitly how the scaling  ratio for the inner and the outermost radius of the Apollonian gasket is identical to that of the scaling ratio between the flux -intervals for  two  successive generations of the butterfly, we note that
from Eq. (\ref{D3}), 
 the ratio of the outer bounding circle and the innermost circles ( See Figs. (\ref{APG}) ) as obtained from
Eq. (\ref{D3}) ) is, 
\begin{equation}
\frac{k_5}{k_4} = \frac{2+\sqrt{3}}{2-\sqrt{3}} = 7-4\sqrt{3} = \zeta^2
\end{equation}

Therefore, the ratio of the bounding to the inner-most circle describe the scaling of the magnetic flux intervals of the butterfly.  See Fig. (\ref{APG}).\\

\begin{equation}
\frac{k_4(+)}{k_4(-)} = \frac{2+\sqrt{3}}{2-\sqrt{3}}
\end{equation}

In the case of butterfly, we have,

\begin{equation}
\sqrt{k_c(\pm)} = \sqrt{k_R(l+1)} \pm \sqrt{k_L(l+1)}= (\sqrt{3} \pm 1) \sqrt{k_L(l+1)},
\end{equation}

which gives,

\begin{equation}  
 \frac{k_c(+)}{k_c(-)} =   \frac{k_c(l+1)}{k_c(l)}=\frac{2+\sqrt{3}}{2-\sqrt{3}}
 \end{equation}

As described above, configuration  of circles underlying the butterfly fractal  appears to be a  special case of a general construction involving four mutually tangent circles. However, we note that
if we consider the mirror image of   the horizontal  (base) line  through the tangency points of the Ford circles corresponding to $f_L(l+1), f_R(l+1), f_c(l)$ and  $f_L(l+1), f_R(l+1), f_c(l+1)$
( See Fig. (\ref{A3})) , we obtain configuration involving four mutually tangent circles, each corresponds to non-zero curvature.
This puts  butterfly fractal
closer to the Apollonian gasket.  We finally remark that although the image circles of the horizontal
line do not correspond to butterflies symmetric about $E=0$, their size scale by the same ratio $\zeta$ and may correspond to off-centered
patterns that are beyond the subject of this paper.\\

Figure (\ref{APG}) shows the integral  Apollonian gaskets, exhibiting hierarchical set of integers that describe quantum numbers of the butterfly obtained by zooming the interval $[1/3-2/5]$.
To obtain complete hierarchy,
of topological integers,  we begin with two
sets of Apollonian gaskets with curvatures $(2,2,3,-1)$ and $(8, 9, 9, -4)$, and use the recursion relation (\ref{pqRR}) for the negative curvatures.
We obtain all the Chern numbers as listed in the first row of Table I , in fact all equivalent circles scale by $\zeta^2$ suggests
that the butterfly fractal characterized by the Farey path "LRL" corresponding to the whole set of irrationals $\zeta_{12}$ are described by the
Apollonian gasket. \\

\begin{widetext}

\begin{figure}[htbp]
\includegraphics[height=4in,width=3.5in]{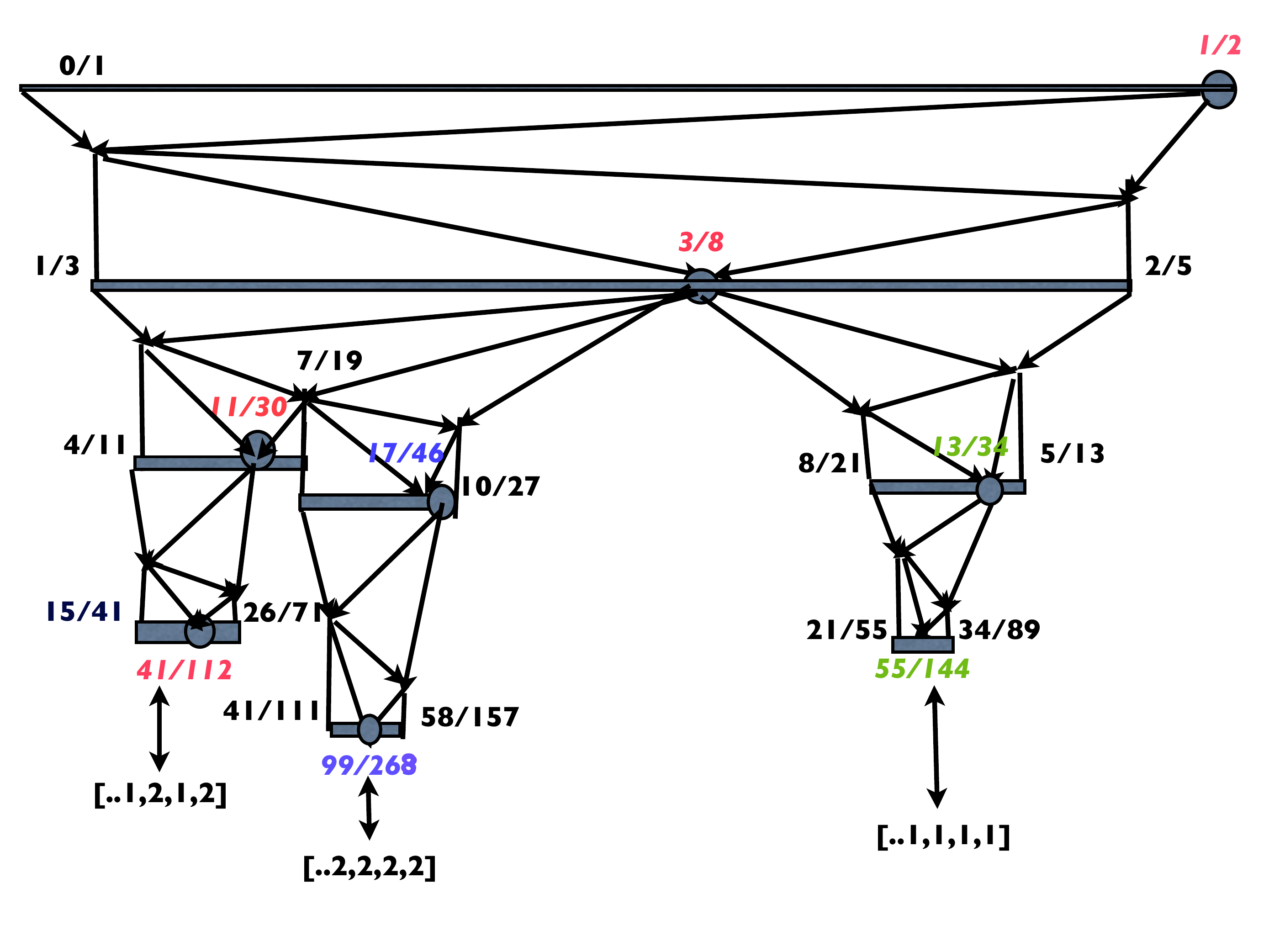}\\
\includegraphics[height=3.5in,width=2in]{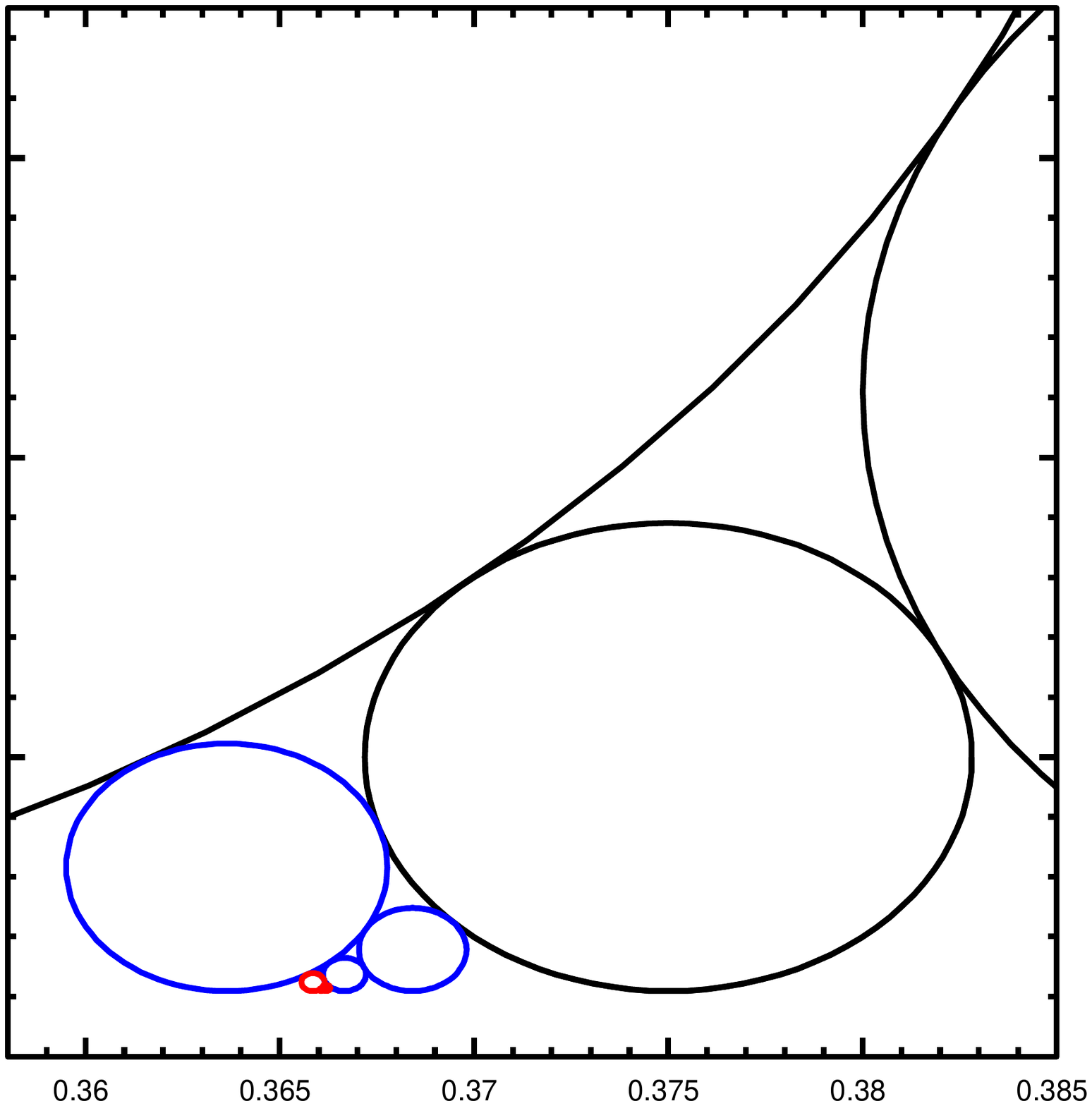}\quad
\includegraphics[height=3.5in,width=2in]{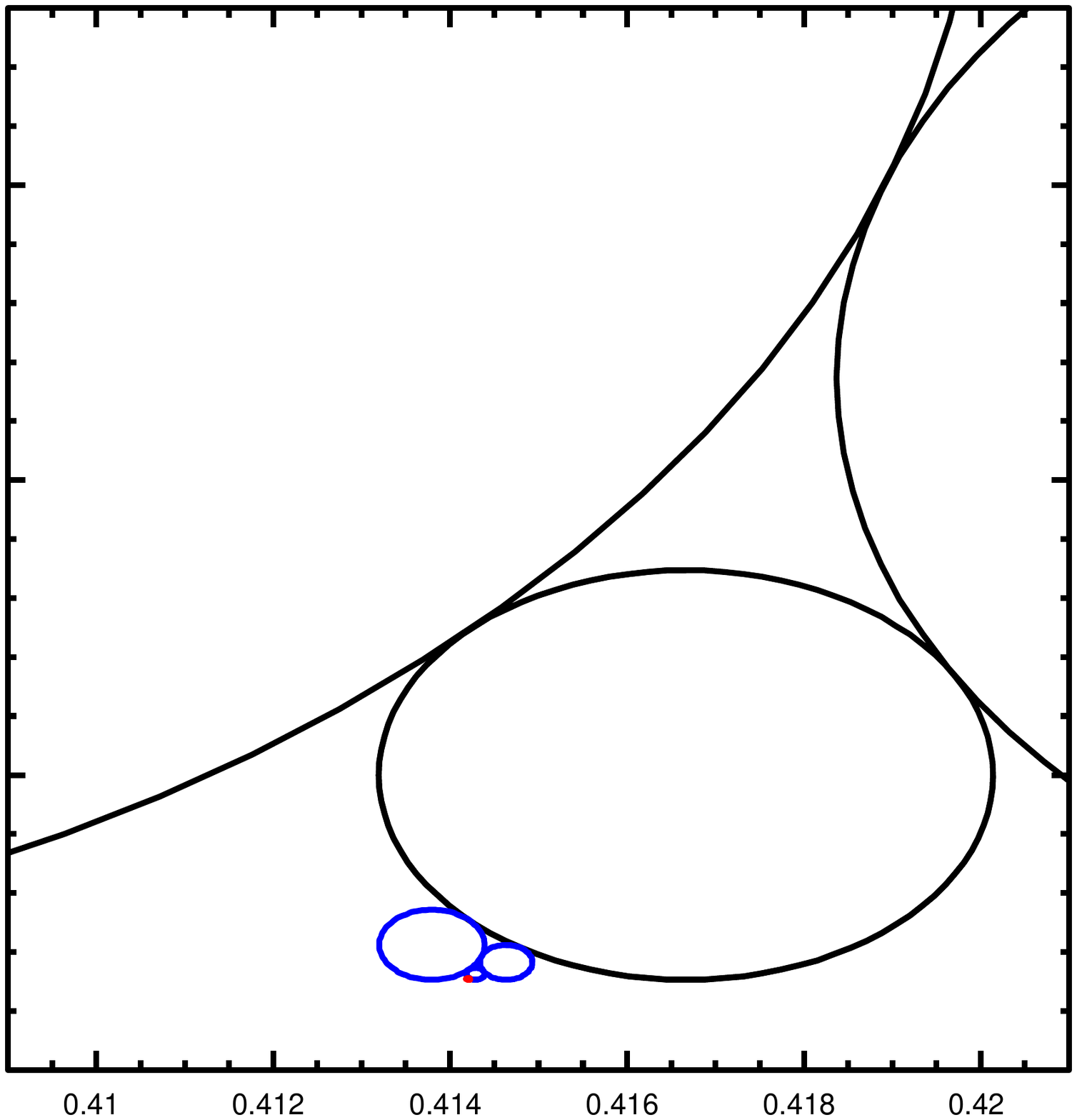}\quad
\includegraphics[height=3.5in,width=2in]{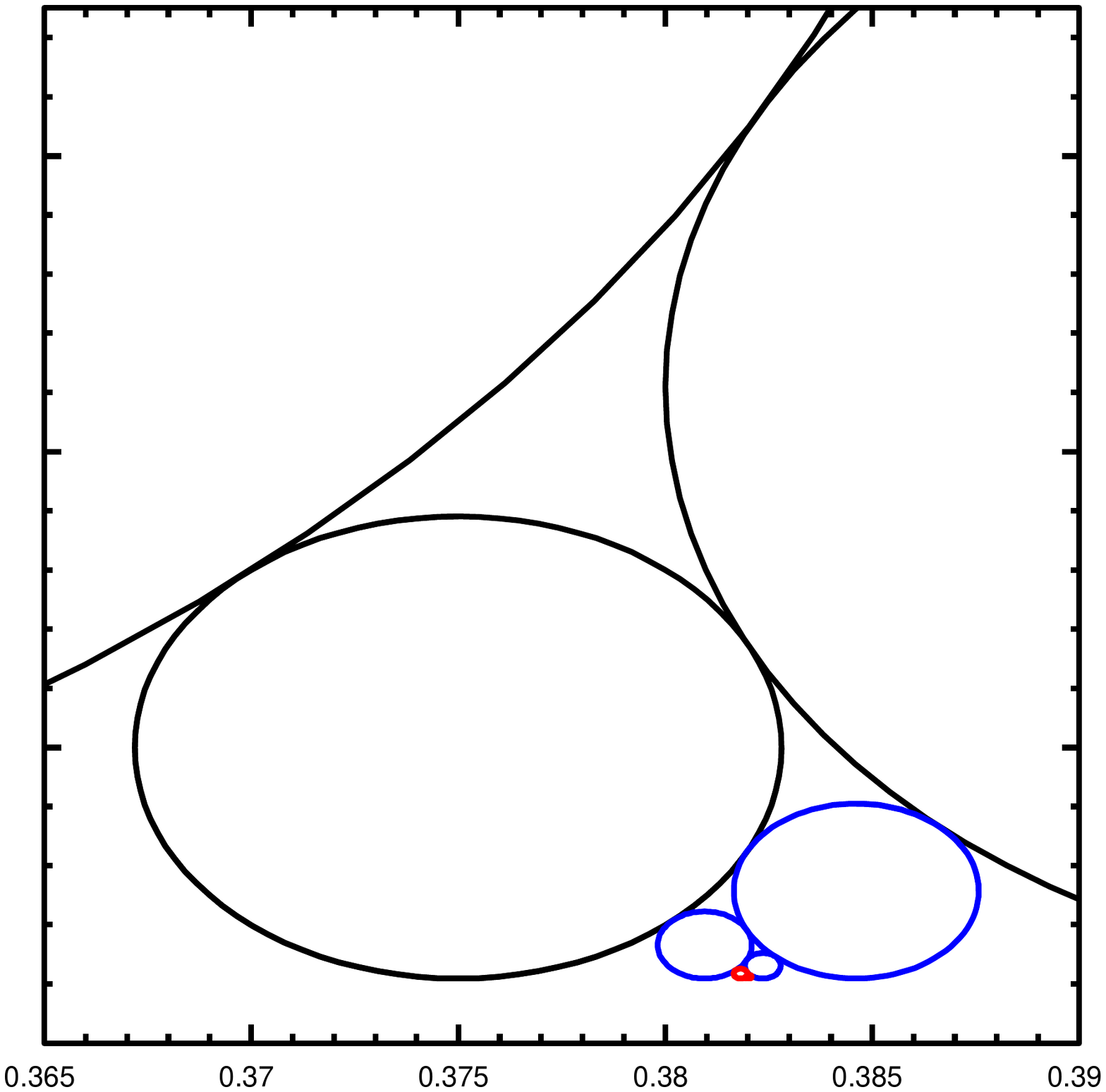}
\leavevmode \caption{  Upper graph is a schematically representation of magnetic flux intervals of the butterfly fractal for three levels of nesting showing $\zeta_{12}$ , $\zeta_{2}$ and $\zeta_1$  hierarchies, all starting with the common level-$1$
sub-interval $1/3-2/5$.  Lower graph shows the corresponding Fords circles ( from left-right ) for $\zeta_{1,2}$, $\zeta_2$ and $\zeta_1$ hierarchies for level-$1$ ( black), level-$2$ (blue) and level-$3$ (red).
This pictorial representation of the center and the boundaries
of the butterfly illustrate  distinctions between the these three hierarchies where the golden-tail butterfly flips between two successive levels. The entire butterfly plot obtained numerically in shown in Fig. (\ref{B3}).}
\label{K3}
\end{figure}

\begin{figure}[htbp]
\includegraphics[height=7in,width=8.5in]{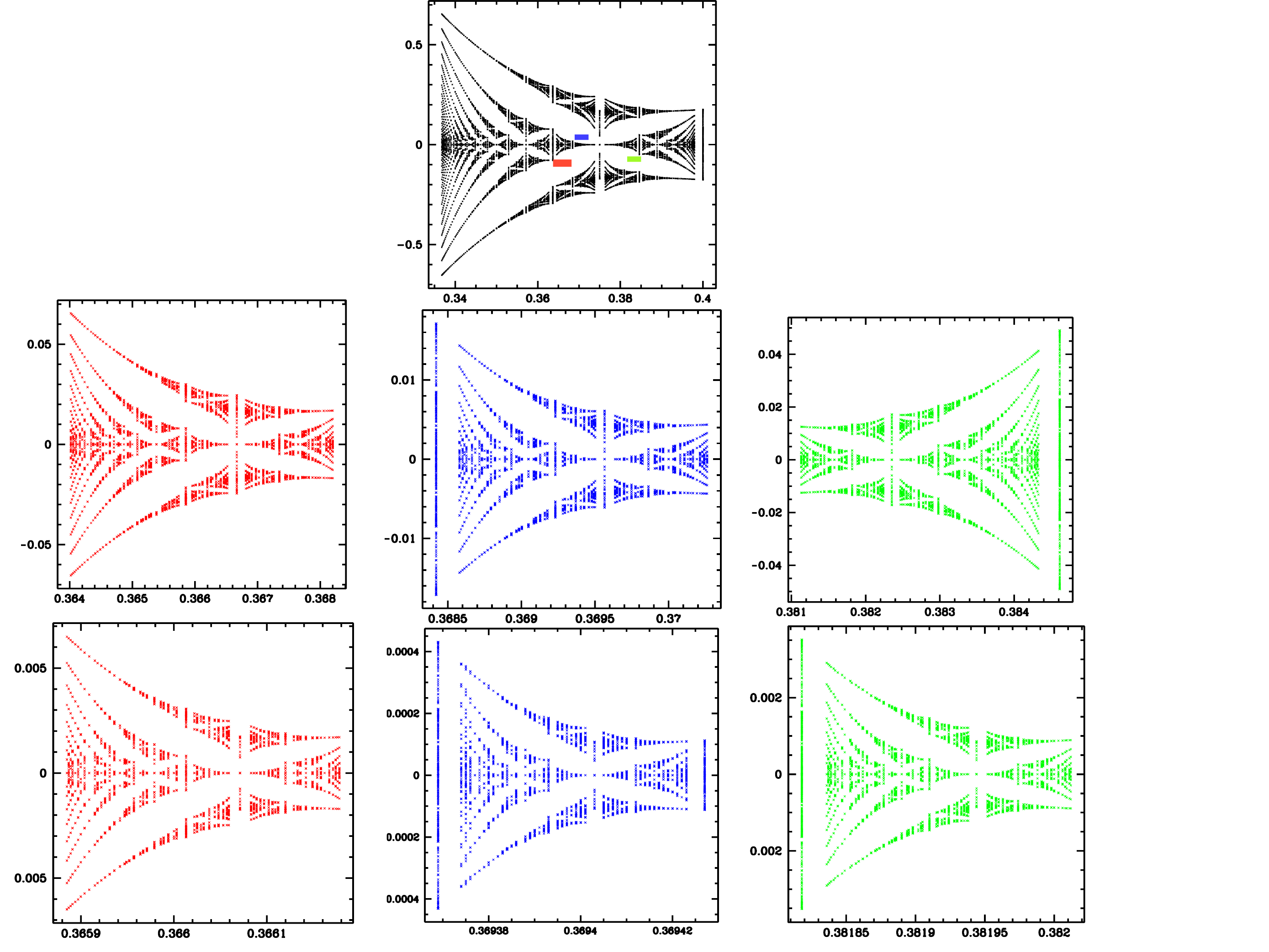}
\leavevmode \caption{ Butterfly plots for three levels for $\zeta_{12}$ ( red), $\zeta_{2}$ ( blue ) and $\zeta_1$ (green)  hierarchies, all starting from the black butterfly at the top. The horizontal lines on the black plot show the appropriate window that is
zoomed and displayed below. We note that golden-mean hierarchy has its butterfly flips between two successive generations.}
\label{B3}
\end{figure}

\begin{table}
\begin{tabular}{| c | c | c | c | c |}
\hline
 $\phi^*_c$\,\, &  Recursion relations for  $q_x, p_x \equiv  s$&  $R_{\phi}$ & $\approx$ $R_E$ &Farey path \\ \hline
 $\zeta_{1,2}$= $ [...1,2,1,2....] $ & $s(l+1)=4s(l)-s(l-1)$ & \,\,  $ 2+\sqrt{3}=3.73205$ & 10 & LRL\\
 \\ \hline
 $\zeta_1$= $ [...1,1,1,1....] $\,\, & $s(l+1)=4s(l)+s(l-1)$ &   $ 2+\sqrt{5}=4.236068$ & 14& LRLRLR\\
 \\ \hline
  $\zeta_{2}$= $ [...2,2,2,2...] $\,\,&   $s(l+1)=6s(l)-s(l-1)$ & $2\sqrt{2}+3=5.82843 $ & 38&  LRRL\\
  \\ \hline
\end{tabular}
\caption{Comparing the scaling ratios $R_{\phi}$  and $R_E$ for various irrational fluxes whose even denominator approximants  form the fixed points of the centers of the butterfly. Each  irrational value
describes a set of numbers with same tail in the continued fraction expansion.} 
\label{table3}
\end{table}

\end{widetext}

\section{ Golden and Silver Mean Hierarchies }

As discussed above, we have investigated the entire Gplot by zooming in the equivalent sets of butterflies and calculating the asymptotic scaling properties of the fixed point butterfly fractal.  In contrast, earlier
studies have explored the butterfly fractal by starting with a fixed irrational number. We  briefly  investigate this line of analysis of the butterfly fractal for the golden and the
silver mean flux values.  We follow the irrational magnetic flux by following a sequence
of its rational approximants with even denominators where
the relation between the boundaries and the center is always given by,
$f_c(l)=f_L(l) \bigoplus f_R(l)$. However, the ordered set  of three rationals $f_L$, $f_c$ and $f_R$ ,  need not belong to the set of rational approximants of the irrational magnetic flux.

For the golden-mean $\gamma_g=(\sqrt{5}-1)/2$ , its
even denominator approximants $(1/2, 3/8,13/34, 55/144, 233/610... )$ form the centers 
of the butterfly at $E=0$ with Chern numbers $(1,4,17,72,305....)$. As shown in
upper part of the Fig. (\ref{K3}), the centers do not form a monotonic sequence and therefore the equivalent set of butterflies correspond to the Farey path $LRLRLR$ or $RLRLRL$.

The silver-mean $\gamma_s = \sqrt{2}-1$  with rational approximants ($1/2, 2/5, 5/12, 12/29, 29/70, 70/169, 169/408....$) result in an silver  hierarchy
with Chern numbers $(1,6,35,204...)$  and correspond to the Farey path $LRRL$ .

Table III compares the three hierarchies  which we will also refer as diamond, silver and golden hierarchy.  Figure (\ref{B3}) shows the three generations of the butterflies
and clearly illustrate the dominance of diamond hierarchy as the scale factors for  both the $\phi$ and the $E$ intervals  are smaller than the corresponding scale factors for the silver and golden cases.

We also note that  unlike diamond hierarchy, silver and golden mean hierarchies  do not map to integral Apollonian gaskets.

\begin{widetext}

\begin{figure}[htbp]
\includegraphics[width = 1.1\linewidth,height=.85\linewidth]{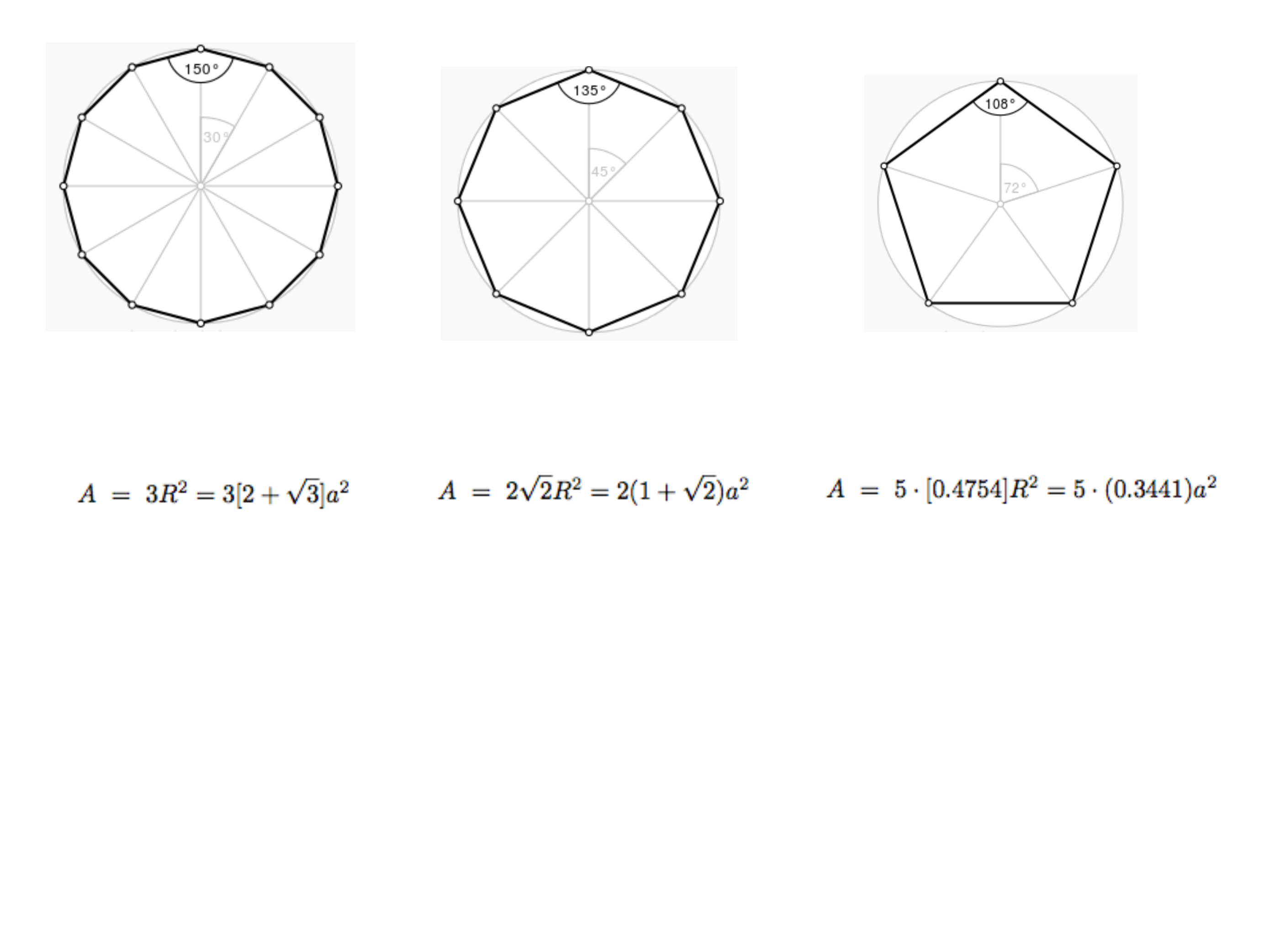}
\leavevmode \caption{ Polygons whose angles relate to the three irrational numbers. Here $R$ is the radius of the circumscribed circle, $a$ is the length of the side of the regular polygon and  $A$ is the area of the polygon.
From left to right ,  the polygons with vortex angles  $30, 45, 72$ degrees  or  equivalently $\frac{2\pi}{12}$, $\frac{2\pi}{8}$ and $\frac{2\pi}{5}$ radian reveal relationship between the three irrationals and the corresponding $12, 8, 5$ fold symmetry.  The area A reveals hidden mathematical simplicity of the diamond mean. }
\label{Poly}
\end{figure}

\end{widetext}

As a final comparison between three irrationals,  we look at the three Polygons, as shown in Fig. (\ref{Poly}) whose angles relate to these irrationals.  Mathematical simplicity of the diamond mean is rather intriguing and it remains to be seen
if the area of dodecagonal being equal to $2 R^2$ ( where $R$ is the radius of the circle ) provides any further insight towards its relationship to $D_3$ symmetry of the Apollonian gasket.

\begin{figure}[htbp]
\includegraphics[width = 1.3\linewidth,height=1.3\linewidth]{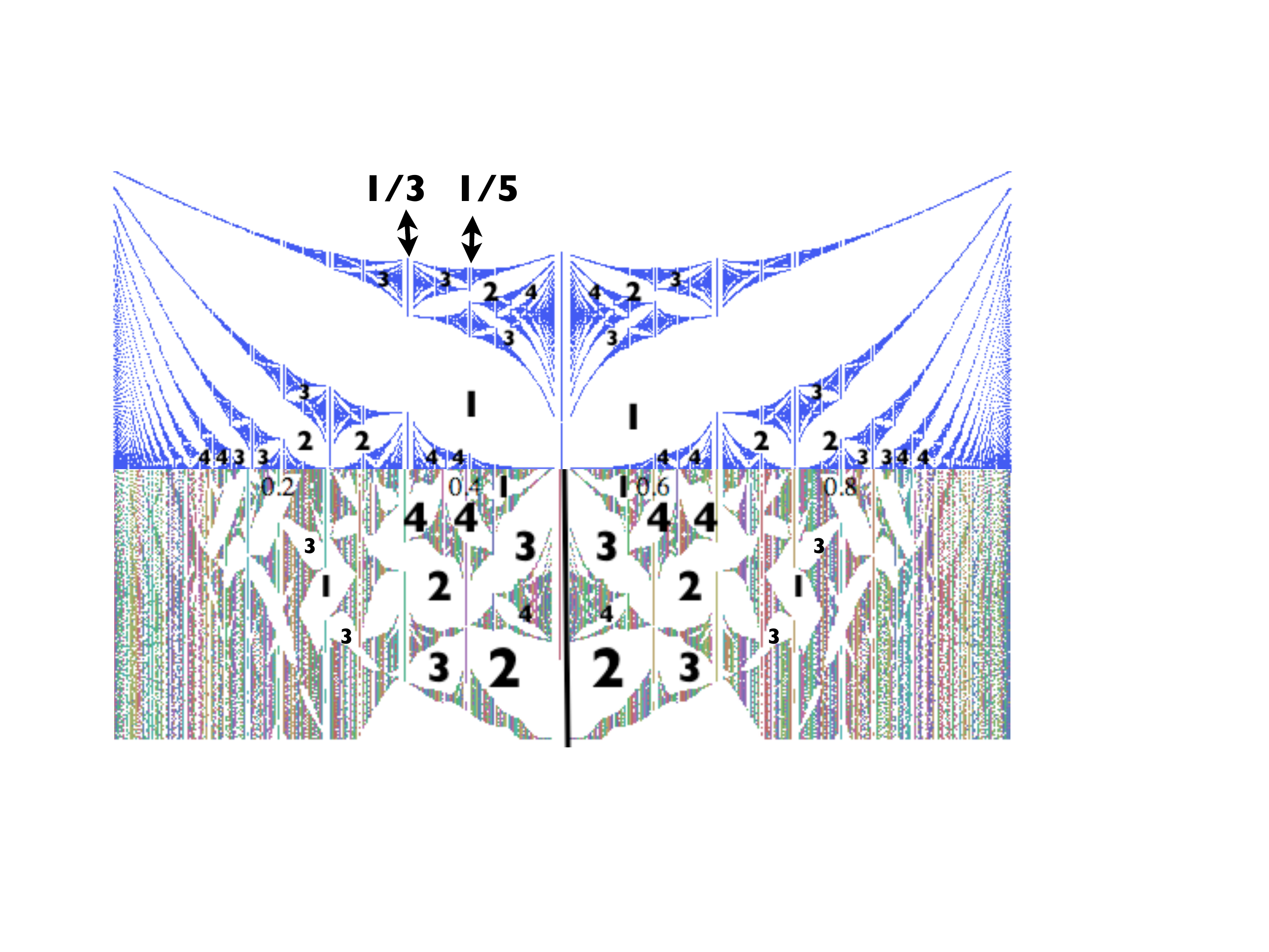}
\leavevmode \caption{(color on line) Upper half of the butterfly spectrum for static ($J_x=J_y$)(upper)
and lower-half of the quasienergy spectrum for driven system with $\bar{J}=.33,\bar{\lambda}=1.1$ (lower).
Dominance of higher Chern states in kicked system is due to
phase transitions where Chern-$1$ state for $\phi=1/3$ 
is transformed
into Chern $-2(=1-3)$ state while Chern ($-2$, $1$) states of $\phi=2/5$ 
evolve into Chern $3(=-2+5)$,$-4(=1-5)$ states.}\label{KickB}
\end{figure}

\section{ Periodic Driving and Gap Amplification}

We next address the question of physical relevance of states of higher topological numbers in view of 
the fact that size of the spectral gaps decreases exponentially with $\sigma$, as confirmed by our
numerical study of the system described by Eq. (\ref{harper}).
We now show that by perturbing such systems, we can induce quantum phase transitions to topological states
with $n > 0$ given by (\ref{DEsol}) with dominant gaps characterized by higher Chern numbers. We study butterfly spectrum for a
periodically kicked quantum Hall system\cite{ML}
where $J_y$ is a periodic function of time $t$ with period-$T$\cite{ML},
\begin{equation}
J_y= \lambda \sum_n \delta(t/T-n)
\end{equation}
The time evolution operator of the system, defined by $|\psi(t)\rangle=U(t)|\psi(0)\rangle$,
has the formal solution $U(t)={\cal T}\exp[-i\int^t_0H(t')dt']$,
where ${\cal T}$ denotes time-ordering and we set $\hbar=1$ throughout.
The discrete
translation symmetry $H(t)=H(t+T)$ leads to a convenient basis
$\{|\phi_\ell\rangle\}$, defined as the eigenmodes of Floquet operator $U(T)$,
\[
U(T)|\phi_\ell\rangle=e^{-i\omega_\ell T}|\phi_\ell\rangle.
\]
We have two independent
driving parameters,
$\bar{J}=J_x T/\hbar$ and $\bar{\lambda}=\lambda T/\hbar$.
For rational flux $\phi=p/q$, 
$U$ is a $q\times q$ matrix with
$q$ quasienergy bands that reduce to the energy bands of the static system as $T \rightarrow 0$.

New topological landscape of the driven system as shown in the Fig. (\ref{KickB}) can be understood by determining the topological states
of flux values corresponding to simple rationals such as $1/3$, $2/5$.  In the Fig. (\ref{KickB}), parameter values correspond to the case
where the Chern-$1$ gap associated with $1/3$ has undergone quantum phase transition to a $n=1$ solution of the DE (Eq. (\ref{DEsol})) and
the Chern-$-2,1$ states of $2/5$ have also undergone transitions to Chern-$3,-4$ state. This almost wipes out
the Chern-$1$ state from the landscape, exposing the topological states of higher Cherns that existed in tiny gaps in the static system.
We note that, topological invariants associated with
irrational flux values can not change under driving and in view of infinite set of irrationals in the vicinity of every rational,
the ordering of the Cherns as we vary the filling factor remains unchanged. 

Gap amplifications of states in periodically driven quantum hall system
may provide a possible pathway to see fractal aspects of
Hofstadter butterfly and engineer states with large Chern numbers experimentally.  Recent experiments with ultracold atoms\cite{Ian2} \cite{Chiding} and shaken
optical lattices\cite{Shake}
 offer a promising means  to realize  the butterfly and its transformation in driven systems.

\section{ Conclusions and Open Questions}

The unveiling of a  dodecagonal quasicrystal\cite{Socolar}, also characterized by integral Apollonian gasket with $D_3$ symmetry
that fully encodes the topological hierarchy of the butterfly fractal is the central result of this paper.   However, the relationship between these two symmetries remain obscure.
We note that these results also apply to other 2D lattices such as graphene in the magnetic field. The associated scaling for topological quantum numbers  is universal, independent of lattice symmetry  and
perhaps indicates result of greater validity and significance. Why dodecagonal quasicrystals emerge as the dominant hierarchy remains an open question. The fact that only these symmetries map to integral
Apollonian gaskets makes the puzzle deeper and more intriguing.
Emergence of hidden symmetries, as energy scale approaches zero is reminiscent of phenomena such as {\it asymptotic freedom} in Quantum Chromodynamics.

Recently, there is renewed interest in quasiperiodic systems\cite{QP, SN,Dana2014} due to
their exotic characteristics that includes their relationship
to topological insulators. Our findings about new symmetries and topological universality will open new avenues in the study of interplay between topology and self-similarity in frustrated systems.

 \appendix
\label{A1}
\section{Geometrical Representation of Fractions: Farey Tree, Ford Circles and Descartes' Theorem}

\begin{figure}[htbp]
\includegraphics[width = 1\linewidth,height=.8\linewidth]{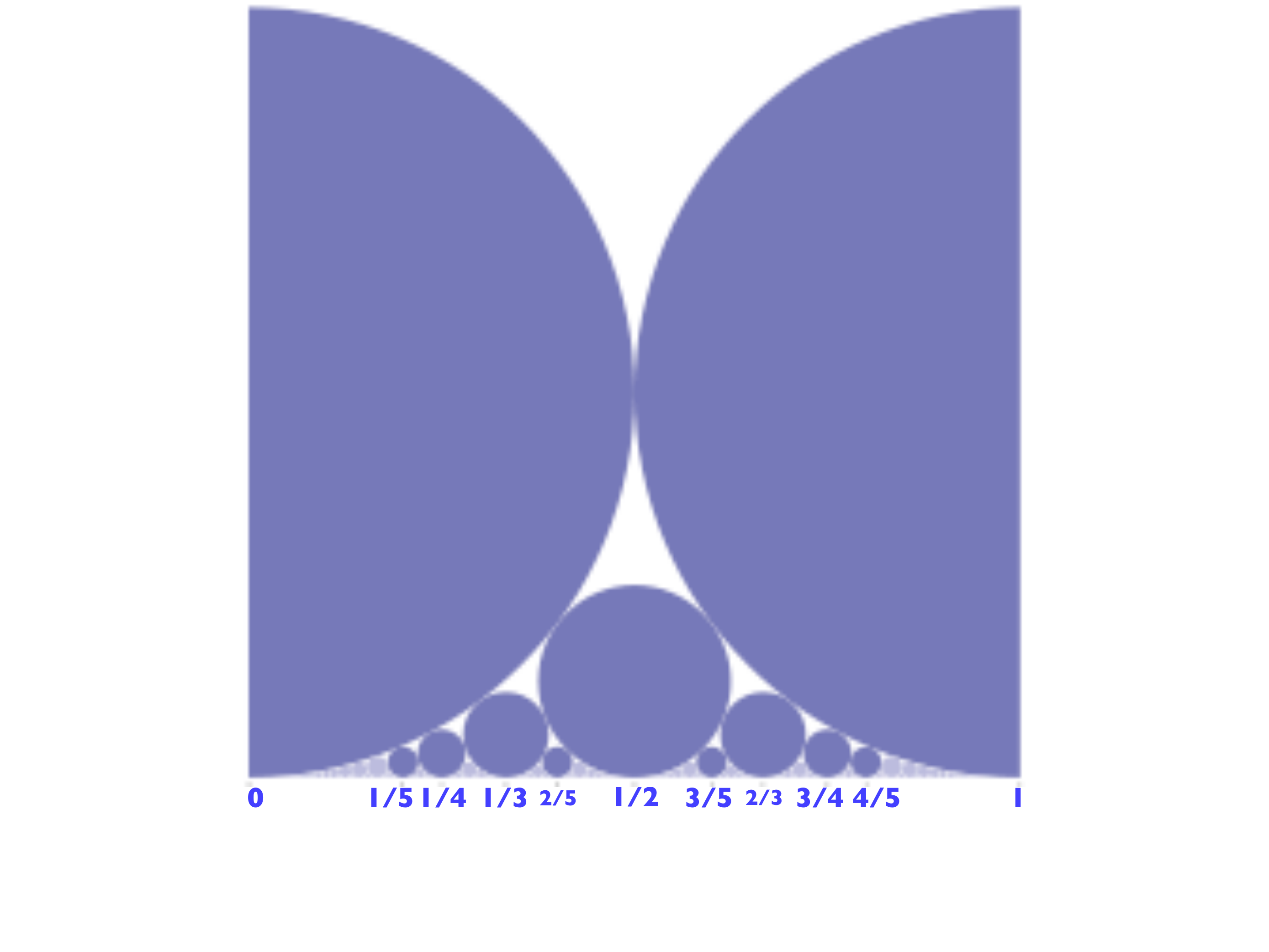}
\leavevmode \caption{ Ford circles provide geometrical representation of Farey sequences. Figure where every fraction $p/q$ is associated with a circle of radius $\frac{1}{2q^2}$,
describes a special case of Descartes's theorem as the straight line is a circle of zero curvature.}
\label{FareyFord}
\end{figure}

\begin{figure}[htbp]
\includegraphics[width =1\linewidth]{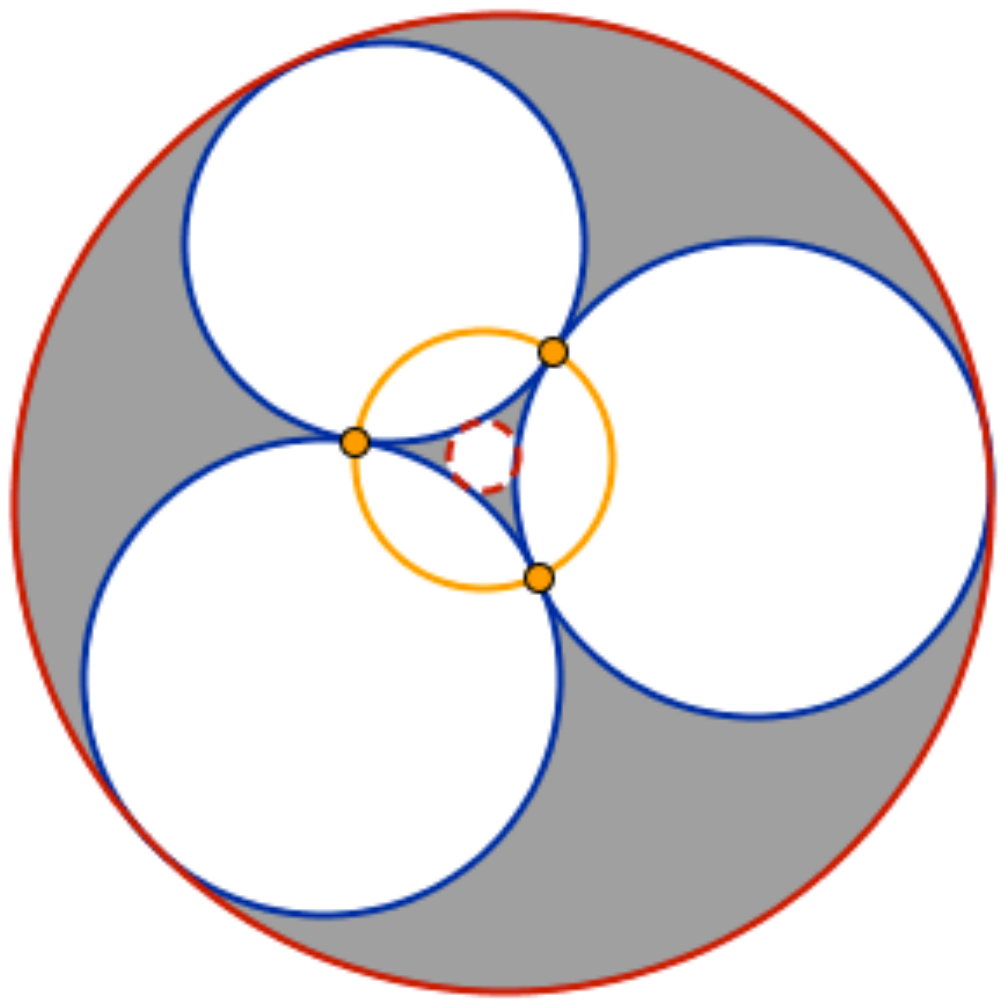}\\
\includegraphics[width =1\linewidth]{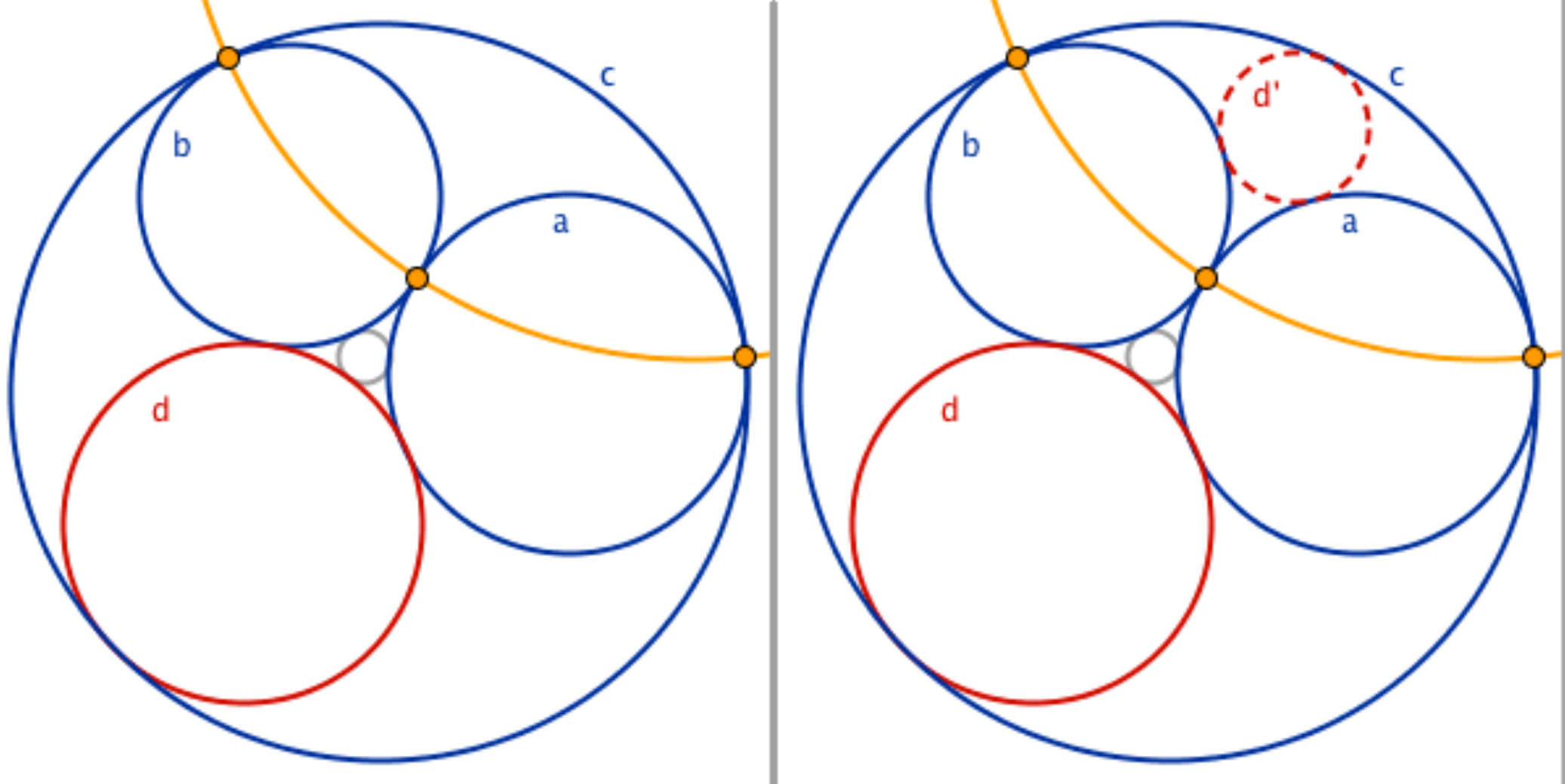}
\leavevmode \caption{ Upper graph illustrates reflection through a circle in a configuration involving four mutually tangent circles. Smallest  ( dashed red) and largest  ( red) circles are mirror image of each other through reflection through the yellow circle, a circle drawn through
the tangency  points of the three circle. The lower graphs illustrates this further, highlighting the preservation of tangency.
The red circle (d)  tangent to the three blue circles, $a, b, c$, and its reflection  in the yellow circle, drawn at the tangency points of the circles $a,b,c$, producing the image ( dotted red) circle labeled $d^{\prime}$. Create a circle through the intersection points of three mutually tangent circles. The three tangent circles are blue in the picture below. }
 \label{tan}
\end{figure}

\underline{ Farey Sequences} of order $n$ is the sequence of completely reduced fractions between $0$ and $1$,
 which have denominators less than or equal to n, arranged in order of increasing size.
We note that two neighboring terms say $\frac{p1}{q1}$ and $\frac{p2}{q2}$ in any Farey sequence satisfy the equation $|q1p2-q2p1|=1$
and  their difference $|\frac{p1}{q1}-\frac{p2}{q2}| =\frac{1}{q1q2}$. 
Of all fractions that are neighbors of $\frac{p}{q}$ ( $ q > 1$ ) {\it only two have denominators less than $q$}. It is this property that associates a unique pair $p_L/q_L$ and $p_R/q_R$ with a fraction
$p_c/q_c$ when $q_c$ is even and $q_L< q_c $ and $q_R < q_c $ are odd\cite{Ford}.
 
\underline{ Ford Circles} ( See Fig. (\ref{FareyFord})) provide geometrical representation of fractions\cite{Ford}.
For every fraction $\frac{p}{q}$ (where $p$ and $q$ are relatively prime) there is a Ford circle, which is the circle with radius $\frac{1}{2q^2}$ and center at $\frac{p}{q}$ and tangent to the base line.
Two Ford circles for different fractions are either disjoint or they are tangent to one another.  In other words, two Ford circles never intersect. 
 If $0 < \frac{p}{q} < 1$, then the Ford circles that are tangent to the ford circle centered at $\frac{p}{q}$, are precisely the Ford circles for fractions that are neighbors of $\frac{p}{q}$ in some Farey sequence. 
 
 Three mutually tangent Ford circles in the Fig. (\ref{FareyFord}), along  with the base line that can be thought of as a circle with infinite radius,  
 are a special case of four mutually tangent circles  such as those shown in Fig. (\ref{A3} R). Relation between  the radii of 
 four such  mutually tangent circles  is given by {\it Descartes's theorem}.\\
   
 \underline{Descartes's theorem} states that 
 if four circles are tangent to each other, and the circles have curvatures  ( inverse of the radius) $k_i$ (for $ i = 1, ..., 4$), a relation
between the curvatures $k_i$ of these circles is given by,

\begin{equation}
(k_1+k_2+k_3+k_4)^2=2(k_1^2+k_2^2+k_3^2+k_4^2).
\label{DT}
\end{equation}
Solving for $k_4$ in terms of $k_i$, $i=1,2,3$ gives,
\begin{equation}
 k_4 = k_1 + k_2 + k_3 \pm 2\sqrt{k_1 k_2 + k_2 k_3 + k_3 k_1}
 \label{Dtheorem}
 \end{equation}
 
The two solutions $\pm$ respectively correspond to the inner ( solid blue circle)  in Fig. (\ref{A3} R ) and the outer bounding circle ( circle with red dots ).  
The consistent solutions of above set of equations require that bounding circle must have negative curvature.  Denoting the curvature of the inner circle as $k_5$,  it follows that

\begin{equation}
k_4 + k_5 = 2(k_1+k_2+k_3)
\label{k5}
\end{equation}

Important consequence of the Eq. (\ref{k5}) is the fact that if $k_i , i= 1-4$ are integers, $k_5$ is also an integer.\\

Patterns obtained by starting with three mutually tangent circles and then recursively inscribing new circles in the curvilinear triangular regions formed between the circles are
Known as the Apollonian gasket,  or
 {\it Curvilinear Sierpinski Gasket}, as the three mutually tangent circles form a triangle in curved space.  An Apollonian gasket describes a packing of circles arising by repeatedly filling the
interstices between four mutually tangent circles with further tangent circles.\\

\underline{ Integral Apollonian Gasket} has all circles whose curvatures are integers. As described above, such a fractal made up integers alone can be constructed if the first four circles have integer curvatures. ( See figs (\ref{APG})).\\

\underline{ Apollonian gaskets with $D_3$ symmetry} is a fractal with $D_3$ symmetry, which corresponds to three reflections along diameters of the bounding circle (spaced $120$ degrees apart), 
along with three-fold rotational symmetry.  Such a gasket can be constructed if the three circles with smallest positive curvature have the same curvature.
 Setting $k_1=k_2=k_3=k$ in Eq. (\ref{Dtheorem}), we obtain 
 \begin{equation}
 k_4({\pm}) = (3 \pm 2 \sqrt{3}) k
 \label{D3}
 \end{equation}
 
 The fact that the ratio $k_4/k$ is an irrational number means that no integral Apollonian circle packing possesses $D_3$ symmetry,
 although many packings come close. 
 As illustrated in the Fig. (\ref{APG}), this symmetry  is restored in the iterative process where the inner circle at level $l$ becomes the outer circle
 at level $l+1$.\\
 
  \underline{Apollonian Gasket-Kaleidoscope} Another remarkable property of the Apollonian gasket  is that
 the whole Apollonian gasket is like a kaleidoscope where the image of the first four circles is reflected again and again through an infinite collection of curved mirrors. 
 
 This is illustrated in Fig. (\ref{tan}) using an operation called {\it inversion},
 a classic tool to understand configurations involving mutually tangent circles,  which is can be thought of as a reflection through a circle. 
 The key feature of the inversion  that maps circles to circles, is that it preserves tangency  as both the circle and its reflected image are tangent to same set of circles as illustrated in Fig. (\ref{tan}).
  
 \bibliography{HBlong} 
\end{document}